\documentclass[aps,prl,reprint,superscriptaddress,amsmath,amssymb]{revtex4-2}

\usepackage{amsthm,mathtools}
\usepackage{graphicx,tikz}
\usetikzlibrary{arrows.meta,shapes.geometric,decorations.pathreplacing}
\usepackage{xcolor}
\definecolor{latnavy}{HTML}{1F3864}
\definecolor{gateblue}{HTML}{2E74B5}
\definecolor{charttwo}{HTML}{2A8C7E}
\definecolor{tgtred}{HTML}{C00000}
\usepackage[colorlinks=true,linkcolor=blue!55!black,citecolor=blue!55!black,urlcolor=blue!55!black]{hyperref}

\theoremstyle{plain}
\newtheorem{theorem}{Theorem}
\newtheorem{lemma}{Lemma}

\newcommand{\cB}{\mathcal B}
\newcommand{\cC}{\mathcal C}
\newcommand{\cH}{\mathcal H}
\newcommand{\cS}{\mathcal S}
\newcommand{\cU}{\mathcal U}
\newcommand{\cV}{\mathcal V}
\newcommand{\Tr}{\operatorname{Tr}}
\newcommand{\eps}{\varepsilon}
\newcommand{\dd}{\mathrm d}
\newcommand{\norm}[1]{\left\lVert #1\right\rVert}
\newcommand{\abs}[1]{\left\lvert #1\right\rvert}
\newcommand{\E}{\mathbb E}
\newcommand{\R}{\mathbb R}
\newcommand{\C}{\mathbb C}
\newcommand{\Ket}[1]{\left\lvert #1\right\rangle}
\newcommand{\Bra}[1]{\left\langle #1\right\rvert}
\newcommand{\PU}{\mathrm{PU}}

\newcommand{\cE}{\mathcal E}
\newcommand{\cR}{\mathcal R}
\newcommand{\id}{\mathrm{id}}
\newcommand{\floor}[1]{\left\lfloor #1\right\rfloor}
\newcommand{\ceil}[1]{\left\lceil #1\right\rceil}
\newcommand{\Set}[1]{\left\{#1\right\}}
\newcommand{\HP}{\mathsf{HP}}
\newcommand{\MSE}{\operatorname{MSE}}
\newcommand{\Arg}{\operatorname{Arg}}

\theoremstyle{definition}
\newtheorem{definition}{Definition}

\theoremstyle{plain}
\newtheorem{smtheorem}[definition]{Theorem}
\newtheorem{smlemma}[definition]{Lemma}
\newtheorem{proposition}[definition]{Proposition}

\begin{document}

\title{Barycentric programming: Classical realization of programs for universal quantum processors}
\author{Yutaka Hashimoto}
\affiliation{Department of Physics, Graduate School of Science, The University of Tokyo, Japan}
\author{Satoshi Yoshida}
\affiliation{Department of Physics, Graduate School of Science, The University of Tokyo, Japan}
\author{Mio Murao}
\affiliation{Department of Physics, Graduate School of Science, The University of Tokyo, Japan}
\affiliation{Trans-Scale Quantum Science Institute, The University of Tokyo, Japan}

\begin{abstract}
A programmable quantum processor implements a target unitary using information stored in a classical or quantum program register.
For arbitrary unitaries in \(\mathrm{SU}(d)\), quantum programs attain the optimal leading-order memory cost, whereas deterministic classical rounding requires twice as much memory for the same diamond-distance error.
We show that this gap disappears when classical randomization is allowed.
We introduce \textit{barycentric programming}, which stores the label of a unitary randomly selected from a finite set, with probabilities chosen so that first-order approximation errors cancel in the averaged unitary channel.
Consequently, the diamond-distance error scales quadratically with the discretization spacing.
For any \(\nu\)-parameter family of unitaries forming a compact, smooth manifold without boundary, this method achieves error \(\eps\) using only \((\nu/2)\log_2(1/\eps)+O(1)\) classical bits.
We also prove a matching lower bound for arbitrary quantum programs, determining the optimal leading-order program size for general unitary families.
Thus, classical randomization achieves the optimal scaling \((\nu/2+o(1))\log_2(1/\eps)\) without requiring coherent quantum memory.
\end{abstract}

\maketitle

\textit{Introduction.}---A universal turing machine performs different tasks on fixed hardware by reading a program that specifies the task. Its quantum counterpart is a programmable quantum processor~\cite{NielsenChuang1997}: a fixed quantum channel that acts jointly on a target system and a program register and applies to the target the unitary encoded in the program state. Nielsen and Chuang showed that no processor with a finite-dimensional program register implements every unitary exactly and deterministically~\cite{NielsenChuang1997}. Processors have therefore been studied in two relaxed forms: probabilistic processors, which are exact but may fail~\cite{NielsenChuang1997,VidalMasanesCirac2002,HilleryBuzekZiman2002,KimCheongLeeLee2001,HilleryZimanBuzek2004,BrazierBuzekKnight2005,SedlakBisioZiman2019}, and approximate processors, which are deterministic but allow an error~\cite{VidalCirac2000,HilleryZimanBuzek2006}, such as port-based teleportation~\cite{IshizakaHiroshima2008}.
Programmable implementation has also been studied for channels~\cite{HilleryZimanBuzek2002}, measurements~\cite{FiurasekDusekFilip2002,DusekBuzek2002,DArianoPerinotti2005,BergouHillery2005,PerezGarcia2006}, and isometries~\cite{YoshidaMiyazakiMurao2026}. See Ref.~\cite{BuzekEtAl2006} for a review. For an approximate processor, the central figure of merit is the program cost, the number of program qubits needed to implement every target within a given error \(\eps\).

Yang, Renner, and Chiribella (YRC)~\cite{YangRennerChiribella2020} determined the optimal program cost for the set of all \(d\)-dimensional unitaries, with the error measured by the diamond norm~\cite{Kitaev1997,WatrousBook2018}: \((\frac{d^2-1}{2}+o(1))\log(1/\eps)\) qubits~\footnote{Here, \(o(1)\) denotes a quantity that tends to zero as \(\eps\to0\), while \(O(1)\) denotes one that remains bounded in magnitude. The dimension \(d\) and the unitary family are held fixed in this limit.}.
In their construction, the target unitary \(U\) acts \(n\) times in parallel on a probe state, and the resulting state \((U^{\otimes n}\otimes I)\Ket\psi\) is compressed by means of its symmetry, as in the compression of identically prepared states~\cite{PleschBuzek2010,YangChiribellaEbler2016,YangChiribellaHayashi2016}. The processor performs a covariant measurement on the compressed state to estimate \(U\) and then applies the estimate. Such an estimation-based strategy is also called a classical strategy~\cite{YoshidaMiyazakiMurao2026}, and in the storage and retrieval of unknown unitaries~\cite{BisioEtAl2010,SedlakBisioZiman2019,SedlakZiman2020,YoshidaMiyazakiMurao2026} it is known to maximize the fidelity for a given number of queries~\cite{BisioEtAl2010,YoshidaMiyazakiMurao2026}. In both cases, however, what is held as the program is the quantum state before the measurement, which must be kept coherent in a quantum memory from its generation to its use. Extracting the estimate by measurement and recording it as classical data instead requires infinite precision, because the estimate is a continuous variable. It has not been known whether a finite classical register can also attain the optimal program size.

\suppressfloats[t]
\begin{figure}[t]
\centering
\pgfdeclareradialshading{mfamily}{\pgfpoint{-14bp}{16bp}}
  {color(0bp)=(white); color(18bp)=(latnavy!5); color(42bp)=(latnavy!18); color(50bp)=(latnavy!30)}
\pgfdeclareradialshading{pububble}{\pgfpoint{0bp}{0bp}}
  {color(0bp)=(white); color(32bp)=(white); color(50bp)=(gray!18)}
\begin{tikzpicture}[
  font=\small,
  x={(1cm,0cm)}, y={(0cm,0.574cm)}, z={(0cm,0.819cm)},
  declare function={
    Rs=1.2; hh=0.26;
    rho(\u,\w)=sqrt(\u*\u+\w*\w)+0.0001;
    cr(\u,\w)=Rs*cos(deg(rho(\u,\w)/Rs));
    sr(\u,\w)=Rs*sin(deg(rho(\u,\w)/Rs))/rho(\u,\w);
    sx(\u,\w,\l,\b)=cr(\u,\w)*sin(\l)*cos(\b)+sr(\u,\w)*(\u*cos(\l)-\w*sin(\l)*sin(\b));
    waist(\x)=0.78-0.30*cos(180*\x/Rs);
    bend(\x)=0.22*sin(180*\x/Rs)+0.06*cos(90*\x/Rs);
    px(\u,\w,\l,\b)=1.08*sx(\u,\w,\l,\b);
    py(\u,\w,\l,\b)=waist(sx(\u,\w,\l,\b))*(-cr(\u,\w)*cos(\l)*cos(\b)+sr(\u,\w)*(\u*sin(\l)+\w*cos(\l)*sin(\b)));
    pz(\u,\w,\l,\b)=waist(sx(\u,\w,\l,\b))*(cr(\u,\w)*sin(\b)+sr(\u,\w)*\w*cos(\b))+bend(sx(\u,\w,\l,\b))/0.819;
  },
  pt/.style={circle,inner sep=0pt,minimum size=1.9pt},
  tgt/.style={circle,fill=tgtred,inner sep=0pt,minimum size=3pt},
  map/.style={-{Stealth[length=2mm]},gray!35!black,line width=0.7pt}
]

\shade[shading=pububble] (0,0) circle[radius=1.6cm];
\draw[gray!35,line width=0.5pt] (0,0) circle[radius=1.6cm];
\node[text=gray!40!black,anchor=south east] at (145:1.55cm) {$\PU(d)$};
\begin{scope}[x=1cm,y=1cm]
  \shadedraw[shading=mfamily,draw=latnavy!40,line width=0.5pt]
    plot[smooth cycle,domain=0:360,samples=81,variable=\t]
      ({1.08*Rs*cos(\t)},{waist(Rs*cos(\t))*Rs*sin(\t)+bend(Rs*cos(\t))});
\end{scope}
\node[text=latnavy,anchor=south west] at (0.85cm,0.85cm) {$M$};

\foreach \l/\b/\c in {-27/2/gateblue,27/2/charttwo}{
  \filldraw[fill=\c,fill opacity=0.22,draw=\c,line width=0.6pt]
    plot[smooth,domain=-3:3,samples=13,variable=\s]
      ({px(\s*hh,-3*hh,\l,\b)},{py(\s*hh,-3*hh,\l,\b)},{pz(\s*hh,-3*hh,\l,\b)})
    -- plot[smooth,domain=-3:3,samples=13,variable=\s]
      ({px(3*hh,\s*hh,\l,\b)},{py(3*hh,\s*hh,\l,\b)},{pz(3*hh,\s*hh,\l,\b)})
    -- plot[smooth,domain=3:-3,samples=13,variable=\s]
      ({px(\s*hh,3*hh,\l,\b)},{py(\s*hh,3*hh,\l,\b)},{pz(\s*hh,3*hh,\l,\b)})
    -- plot[smooth,domain=3:-3,samples=13,variable=\s]
      ({px(-3*hh,\s*hh,\l,\b)},{py(-3*hh,\s*hh,\l,\b)},{pz(-3*hh,\s*hh,\l,\b)}) -- cycle;
  \foreach \i in {-2,...,2}{
    \draw[\c!60,line width=0.35pt]
      plot[smooth,domain=-3:3,samples=13,variable=\s]
      ({px(\i*hh,\s*hh,\l,\b)},{py(\i*hh,\s*hh,\l,\b)},{pz(\i*hh,\s*hh,\l,\b)});
    \draw[\c!60,line width=0.35pt]
      plot[smooth,domain=-3:3,samples=13,variable=\s]
      ({px(\s*hh,\i*hh,\l,\b)},{py(\s*hh,\i*hh,\l,\b)},{pz(\s*hh,\i*hh,\l,\b)});
  }
  \foreach \i in {-3,...,3}\foreach \j in {-3,...,3}
    \node[pt,fill=\c] at ({px(\i*hh,\j*hh,\l,\b)},{py(\i*hh,\j*hh,\l,\b)},{pz(\i*hh,\j*hh,\l,\b)}) {};
}

\def\l{-27}\def\b{2}
\draw[gateblue,line width=0.9pt]
  ({px(-2*hh,0,\l,\b)},{py(-2*hh,0,\l,\b)},{pz(-2*hh,0,\l,\b)})
  -- ({px(-1*hh,0,\l,\b)},{py(-1*hh,0,\l,\b)},{pz(-1*hh,0,\l,\b)})
  -- ({px(-1*hh,1*hh,\l,\b)},{py(-1*hh,1*hh,\l,\b)},{pz(-1*hh,1*hh,\l,\b)})
  -- ({px(-2*hh,1*hh,\l,\b)},{py(-2*hh,1*hh,\l,\b)},{pz(-2*hh,1*hh,\l,\b)}) -- cycle;
\node[tgt] (tM) at ({px(-1.62*hh,0.4*hh,\l,\b)},{py(-1.62*hh,0.4*hh,\l,\b)},{pz(-1.62*hh,0.4*hh,\l,\b)}) {};
\draw[gray!60,line width=0.45pt]
  (tM) -- (-1.85cm,-0.2cm) node[left,inner sep=1pt,text=tgtred] {$F_1(t)$};
\node[pt,fill=gateblue,minimum size=3.2pt] (Vx)
  at ({px(-3*hh,-1*hh,\l,\b)},{py(-3*hh,-1*hh,\l,\b)},{pz(-3*hh,-1*hh,\l,\b)}) {};
\draw[gray!60,line width=0.45pt]
  (Vx) -- (-1.85cm,-0.62cm) node[left,inner sep=1pt,text=gateblue] {$\cV_x$};
\coordinate (chartone) at ({px(0,-2*hh,-27,2)},{py(0,-2*hh,-27,2)},{pz(0,-2*hh,-27,2)});
\coordinate (charttwo) at ({px(0,-2*hh,27,2)},{py(0,-2*hh,27,2)},{pz(0,-2*hh,27,2)});

\foreach \sx/\sy/\xx/\xy/\yx/\c/\a in {-2.55/-2.35/0.906/0.243/-0.423/gateblue/1,
                                      1.19/-1.99/0.906/-0.243/0.423/charttwo/2}{
  \begin{scope}[shift={(\sx cm,\sy cm)},x={(\xx cm,\xy cm)},y={(\yx cm,0.520cm)}]
    \fill[\c!7,draw=\c!75,line width=0.5pt] (0,0) -- (1.5,0) -- (1.5,1.5) -- (0,1.5) -- cycle;
    \foreach \i in {1,...,5}{
      \draw[\c!40,line width=0.35pt] (\i*0.25,0) -- (\i*0.25,1.5);
      \draw[\c!40,line width=0.35pt] (0,\i*0.25) -- (1.5,\i*0.25);
    }
    \foreach \i in {0,...,6}\foreach \j in {0,...,6}
      \node[pt,fill=\c] at (\i*0.25,\j*0.25) {};
  \end{scope}
}
\begin{scope}[shift={(-2.55cm,-2.35cm)},x={(0.906cm,0.243cm)},y={(-0.423cm,0.520cm)}]
  \draw[gateblue,line width=0.9pt] (0.25,0.75) -- (0.5,0.75) -- (0.5,1.0) -- (0.25,1.0) -- cycle;
  \node[tgt] (tB) at (0.25+0.38*0.25,0.75+0.4*0.25) {};
  \node[pt,fill=gateblue,minimum size=3.2pt] (xB) at (0,0.5) {};
\end{scope}
\node[anchor=east,inner sep=2pt,text=gateblue] at (xB.west) {$x$};
\draw[gray!60,line width=0.45pt]
  (tB) -- (-3.12cm,-1.78cm) node[left,inner sep=1pt,text=tgtred] {$t$};
\node[anchor=north east,inner sep=1pt] at (-2.6cm,-2.38cm) {$B_1$};
\node[anchor=north west,inner sep=1pt] at (2.6cm,-2.38cm) {$B_2$};
\draw[map] (-1.55cm,-1.45cm) to[out=60,in=210] node[left,pos=0.5,inner sep=2pt] {$F_1$}
  (chartone);
\draw[map] (1.55cm,-1.45cm) to[out=120,in=-30] node[right,pos=0.5,inner sep=2pt] {$F_2$}
  (charttwo);
\end{tikzpicture}
\caption{Lattice of labels in barycentric programming. The family \(M\subset\PU(d)\) is covered by finitely many overlapping charts \(F_\alpha\) defined on rectangles \(B_\alpha\subset\R^\nu\), so that each point of \(M\) can be written as \(F_\alpha(t)\) with coordinates \(t\in B_\alpha\). The points \(x\) of a lattice with spacing \(h=\Theta(\sqrt\eps)\) in \(B_\alpha\) serve as labels \((\alpha,x)\), and each label carries the unitary \(\cV_x=F_\alpha(x)\). A target (red) is rounded at random to the vertices of the cell containing its coordinates \(t\).}
\label{fig:lattice}
\end{figure}
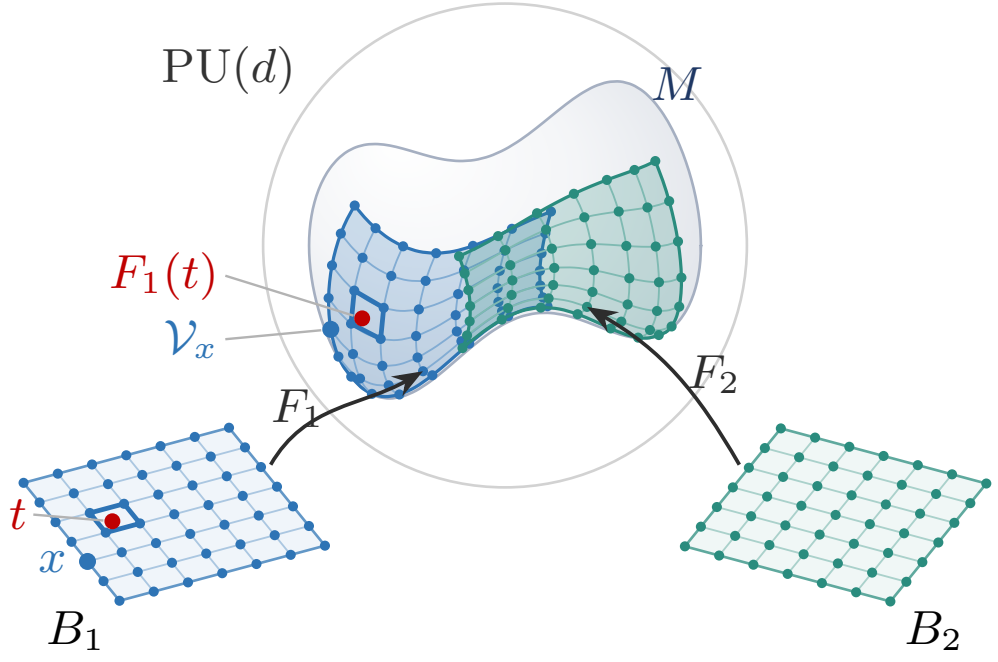

\begin{figure*}[t]
\centering
\newcommand{\roundingcell}[5]{%
  \begin{scope}[shift={(#1,#2)}]
    \draw[gray!40,line width=0.4pt] (-0.72,-0.5) -- (0.72,-0.5) (-0.72,0.5) -- (0.72,0.5)
      (-0.5,-0.72) -- (-0.5,0.72) (0.5,-0.72) -- (0.5,0.72);
    \draw[latnavy,line width=0.8pt] (-0.5,-0.5) rectangle (0.5,0.5);
    \foreach \vx/\vy in {-0.5/-0.5,0.5/-0.5,-0.5/0.5,0.5/0.5}
      \fill[latnavy] (\vx,\vy) circle[radius=2pt];
    \node[gate,anchor=north east] at (-0.53,-0.53) {$\cV_1$};
    \node[gate,anchor=north west] at (0.53,-0.53) {$\cV_2$};
    \node[gate,anchor=south east] at (-0.53,0.53) {$\cV_3$};
    \node[gate,anchor=south west] at (0.53,0.53) {$\cV_4$};
    \fill[tgtred] (-0.5+1.0*#3,-0.5+1.0*#4) coordinate (est) circle[radius=1.6pt];
    \node[anchor=west] at (1.05,0)
      {$\textcolor{tgtred}{\widehat\cU_{#5}}\approx\displaystyle\sum_jp_j^{(#5)}\,\textcolor{gateblue}{\cV_j}$};
  \end{scope}}
\begin{tikzpicture}[
  font=\small,
  wire/.style={gray!45!black,line width=0.6pt},
  slot/.style={draw=gray!50!black,fill=gray!22,line width=0.5pt,minimum width=0.46cm,
    minimum height=0.4cm,inner sep=0pt},
  branch/.style={-{Stealth[length=1.8mm]},gray!35!black,line width=0.6pt},
  gate/.style={text=gateblue,inner sep=1pt,font=\scriptsize},
  estlabel/.style={text=tgtred,inner sep=2pt,font=\scriptsize}
]

\begin{scope}[shift={(0,-0.5)}]
  \filldraw[fill=gray!8,draw=gray!45!black,line width=0.6pt,rounded corners=0.8pt]
    (0,0) -- (4.78,0) -- (4.78,1) -- (4.4,1) -- (4.4,0.28) -- (3.68,0.28) -- (3.68,1)
    -- (3.3,1) -- (3.3,0.28) -- (2.58,0.28) -- (2.58,1) -- (2.2,1) -- (2.2,0.28)
    -- (1.48,0.28) -- (1.48,1) -- (1.1,1) -- (1.1,0.28) -- (0.38,0.28) -- (0.38,1)
    -- (0,1) -- cycle;
  \foreach \c in {0.74,1.84,4.04}{
    \draw[wire] (\c-0.36,0.64) -- (\c+0.36,0.64);
    \node[slot] at (\c,0.64) {$\cU$};
  }
  \draw[wire] (2.58,0.64) -- (2.7,0.64) (3.18,0.64) -- (3.3,0.64);
  \node at (2.94,0.64) {$\cdots$};
  \draw[decorate,decoration={brace,amplitude=4pt},gray!40!black,line width=0.5pt]
    (0.5,1.12) -- node[above=4pt,text=black] {$\Theta(\eps^{-1/2})$ uses of $\cU$} (4.28,1.12);
  \draw[wire] (4.78,0.5) -- (5.12,0.5);
  \filldraw[fill=gray!8,draw=gray!45!black,line width=0.6pt,rounded corners=0.8pt]
    (5.12,0.2) rectangle (5.72,0.8);
  \draw[gray!30!black,line width=0.5pt] (5.24,0.38) arc[start angle=160,end angle=20,radius=0.19];
  \draw[gray!30!black,line width=0.5pt,-{Stealth[length=1mm]}] (5.42,0.34) -- (5.56,0.62);
\end{scope}

\roundingcell{7.6}{1.15}{0.25}{0.75}{1}
\node[estlabel,anchor=west] at (est) {$\widehat\cU_1$};
\node[anchor=south,inner sep=1.5pt,font=\scriptsize,text=gray!30!black] at (7.6,1.65) {$\Theta(\sqrt\eps)$};
\roundingcell{7.6}{-0.75}{0.7}{0.3}{2}
\node[estlabel,anchor=east] at (est) {$\widehat\cU_2$};
\node at (7.6,-1.85) {$\vdots$};
\node at (10.0,-1.85) {$\vdots$};
\draw[branch] (5.85,0) -- node[above left=-1pt,pos=0.55] {$q_1$} (6.75,1.0);
\draw[branch] (5.85,0) -- node[above,pos=0.45] {$q_2$} (6.75,-0.65);
\draw[branch] (5.85,0) -- (6.75,-1.75);

\draw[decorate,decoration={brace,amplitude=6pt},gray!40!black,line width=0.5pt]
  (11.4,2.0) -- (11.4,-2.1);
\node[anchor=west] at (11.7,0.55)
  {$\cU\approx\displaystyle\sum_iq_i\,\textcolor{tgtred}{\widehat\cU_i}
     =\sum_{ij}q_i\,p_j^{(i)}\,\textcolor{gateblue}{\cV_j}$};
\node[anchor=west] at (11.7,-0.85)
  {$\sigma=\displaystyle\sum_{ij}q_i\,p_j^{(i)}\,\textcolor{gateblue}{\Ket j\!\Bra j}$};
\end{tikzpicture}
\caption{Generation of a classical program from an unknown unitary \(\cU\). The estimation stage uses \(n=\Theta(\eps^{-1/2})\) queries to the unitary (left). Each estimate \(\widehat\cU_i\), obtained with probability \(q_i\), is rounded unbiasedly to the unitaries \(\cV_j\) at the vertices of the enclosing lattice cell of width \(\Theta(\sqrt\eps)\) (middle). The two stages of randomness generate the classical program state \(\sigma\), which stores label \(j\) with probability \(\sum_i q_i p_j^{(i)}\). The processor reads this label and applies \(\cV_j\), so its average channel approximates \(\cU\) within \(\eps\) (right). For illustration, the estimate is drawn as taking discrete values \(\widehat\cU_i\). In general it is a continuous random variable, and the sums over \(i\) become integrals over its distribution.}
\label{fig:overview}
\end{figure*}
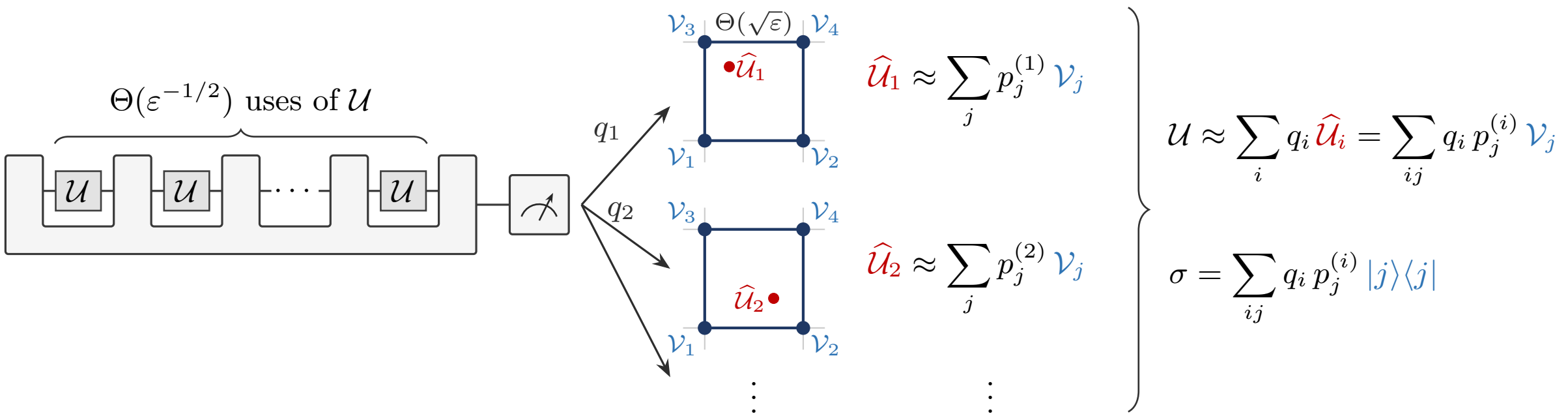

A simple classical strategy labels a nearby unitary from a finite set, as in the truncated-angle example of Ref.~\cite{VidalMasanesCirac2002} and the \(\eps\)-net construction of Ref.~\cite{KubickiPalazuelosPerezGarcia2019}. Covering all \(d\)-dimensional unitaries to diamond-distance accuracy \(\eps\) requires \((d^2-1)\log(1/\eps)+O(1)\) bits~\cite{HarrowRechtChuang2002,OszmaniecSawickiHorodecki2022}, twice the optimal quantum cost to leading order. On this basis, YRC concluded that quantum programs offer a quadratic advantage in program dimension over classical programs~\cite{YangRennerChiribella2020}. YRC conjectured that, for a family of unitaries described by \(\nu\) real parameters, the optimal cost is \((\frac\nu2+o(1))\log(1/\eps)\)~\cite{YangRennerChiribella2020}. Specific families have been analyzed, including covariant channels~\cite{GschwendtnerBluhmWinter2021}, Gaussian channels~\cite{GschwendtnerWinter2021}, reflections~\cite{SchouteEtAl2024}, and two-outcome projective meaurements~\cite{yoshida2025universal}. For the reflections \(I-2\Ket\psi\!\Bra\psi\), Schoute et al.\ studied how to implement the reflection from copies of \(\psi\), which yields a quantum program of cost \((d-1)\log(1/\eps)+O(1)\), and derived a matching lower bound under two conjectures~\cite{SchouteEtAl2024}. He and Yang showed that the cost of a general family with a Lipschitz parametrization is at least \((\frac{\nu-1}{2}-o(1))\log(1/\eps)\), and their net construction with orthogonal label states, which is therefore classical, attains \(\nu\log(1/\eps)+O(1)\) for a fixed family~\cite{HeYang2026}; the leading coefficients of these bounds differ by a factor of more than two.

\begin{table*}[t]
\caption{\label{tab:comparison}Achievable program costs for implementing a unitary within error \(\eps\), as \(\eps\to0\) with \(d\) or the family \(M\) fixed. The upper bound of this paper is tight in the leading term (Theorem~\ref{thm:main}).}
\begin{ruledtabular}
\begin{tabular}{lllll}
Family & Construction & Program & Achievable cost (upper bound) & Generation\\
\hline
\(\PU(d)\) & \(\eps\)-net~\cite{KubickiPalazuelosPerezGarcia2019} & Classical & \(d^2\log(1/\eps)+O(1)\) & Classical description assumed\\
 & YRC~\cite{YangRennerChiribella2020} & Quantum & \(\frac{d^2-1}{2}\log(1/\eps)+O(1)\) & \(O(d^2\eps^{-1/2})\) queries\\
 & This paper & Classical & \(\frac{d^2-1}{2}\log(1/\eps)+O(1)\) & \(O(d^2\eps^{-1/2})\) queries\\
\hline
Reflections & Schoute et al.~\cite{SchouteEtAl2024} & Quantum & \((d-1)\log(1/\eps)+O(1)\) & \(O(1/\eps)\) copies of \(\psi\)\\
 & This paper & Classical & \((d-1)\log(1/\eps)+O(1)\) & \(O(\eps^{-1/2})\) queries\\
\hline
\(\nu\)-parameter family & He and Yang~\cite{HeYang2026} & Classical & \(\nu\log(1/\eps)+O(1)\) & Classical description assumed\\
 & This paper & Classical & \(\frac\nu2\log(1/\eps)+O(1)\) & \(O(\eps^{-1/2})\) queries\\
\end{tabular}
\end{ruledtabular}
\end{table*}

In this Letter, we show that a finite classical program suffices to attain the optimal program size. To this end, we introduce \textit{barycentric programming}, which constructs a classical program by storing the label of a unitary randomly selected from a finite set. For every family of unitaries described smoothly by \(\nu\) real parameters, barycentric programming attains the optimal program cost \((\frac\nu2+o(1))\log(1/\eps)\) with a label from a set of \(O(\eps^{-\nu/2})\) labels. The processor reads the label and applies the corresponding unitary. Together with a matching lower bound, this proves the YRC conjecture. The same program can be generated from \(O(\eps^{-1/2})\) queries to the target given as an unknown black box. Since the program is a classical label, it can be held in classical memory instead of quantum memory and transmitted over a classical channel at the optimal cost. Table~\ref{tab:comparison} compares these results with existing constructions.

We first cover the unitary family with finitely many overlapping coordinate charts and place a lattice in each coordinate domain. Each lattice point serves as a classical label for the unitary specified by the corresponding chart (Fig.~\ref{fig:lattice}). We then combine this lattice construction with randomized unitary synthesis. By linearity in the program state, the processor implements a weighted average, or barycenter, of the unitary channels associated with the labels~\cite{BanchiEtAl2020}. Only this average needs to approximate the target. Unbiased randomized rounding cancels the first-order error, making the error quadratic in the lattice spacing and allowing a mesh of \(\Theta(\sqrt\eps)\)~\cite{Campbell2017,Hastings2017,AkibueKatoTani2021,AkibueKatoTani2024,KliuchnikovEtAl2023}. The same rounding has also been studied for classical random variables as dual quantization~\cite{PagesWilbertz2012}. For an unknown target, a Heisenberg-limited estimator~\cite{HaahKothariODonnellTang2023} and a pullback of the estimate onto the family precede the rounding (Fig.~\ref{fig:overview}). Our construction and matching lower bound apply to general smooth families of unitaries, without requiring the family to form a group.

\textit{Setting.}---All Hilbert spaces are finite dimensional, \(\cH\simeq\C^d\), and logarithms are to base 2. We write \(\PU(d)=\mathrm{U}(d)/\mathrm{U}(1)\) for the projective unitary group and identify each equivalence class \(u=[U]\) with the unitary channel \(\cU_u(\rho)=U\rho U^\dagger\), where \([U]\) denotes \(U\) up to a global phase.
We consider a family of unitaries described locally by \(\nu\) real parameters, where \(\nu\ge1\).
We represent this family as a \(\nu\)-dimensional submanifold \(M\subset\PU(d)\) and write \(\cS=\{\cU_u\}_{u\in M}\) for the corresponding family of channels.
We work in \(\PU(d)\) instead of \(\mathrm{SU}(d)\) so that distinct points of \(M\) represent distinct channels.
To quantify the approximation error, we take the diamond distance defined by ${1\over 2}\|\Phi-\Lambda\|_\diamond$ using the diamond norm $\|\cdot\|_\diamond$~\cite{Kitaev1997}. For this family, an \(\eps\)-processor consists of a \(u\)-independent channel \(\cC:\cB(\cH\otimes\cH_P)\to\cB(\cH)\) and program states \(\sigma_u\) on \(\cH_P\) satisfying
\begin{equation}
  \tfrac12\norm{\cC(\,\cdot\,\otimes\sigma_u)-\cU_u}_\diamond\le\eps
  \qquad(u\in M).
  \label{eq:processor}
\end{equation}
The optimal program cost \(c_P^*(\eps;\cS)\) is the infimum of \(\log\dim\cH_P\) over all \(\eps\)-processors. The program is called classical if all \(\sigma_u\) are diagonal in a \(u\)-independent basis \(\{\Ket x\}_{x\in\Gamma}\), that is, if it is a random label \(x\in\Gamma\) with distribution \(q(x\mid u)=\Bra x\sigma_u\Ket x\). When the target is an unknown black box, the program is generated from \(n\) queries if \(\sigma_u\) is the output of a \(u\)-independent adaptive strategy, or quantum comb~\cite{ChiribellaDArianoPerinotti2008architecture,ChiribellaDArianoPerinotti2009networks,GutoskiWatrous2007}, with at most \(n\) slots filled by \(\cU_u\). Auxiliary qubits, intermediate measurements, and classical feed-forward are allowed.

We further assume that \(M\) is compact, without boundary, and of class \(C^3\). Examples are \(\PU(d)\) itself (\(\nu=d^2-1\)), the qubit phase gates (\(\nu=1\)), and the reflections \(I-2\Ket\psi\!\Bra\psi\) (\(\nu=2(d-1)\)).

\begin{theorem}
\label{thm:main}
Let \(d\ge2\). There is a constant \(C<\infty\) depending only on \(M\) such that, for all sufficiently small \(\eps>0\),
\begin{equation}
  \frac\nu2\log\frac1\eps
  -\frac{3\nu}2\log\log\frac1\eps-C
  \le c_P^*(\eps;\cS)
  \le\frac\nu2\log\frac1\eps+C .
  \label{eq:main}
\end{equation}
The upper bound is attained by a classical program over \(O(\eps^{-\nu/2})\) labels, which can be generated using \(O(\eps^{-1/2})\) queries to the unknown channel \(\cU_u\).
\end{theorem}

For \(M=\PU(d)\), Theorem~\ref{thm:main} reproduces the leading term of YRC, and the number of queries, \(O(d^2\eps^{-1/2})\) (see the proof of Lemma~\ref{lem:haah} in the SM), coincides with that of their quantum program.
Since the lower bound holds for arbitrary quantum programs, quantum programs offer no advantage in the leading term. We describe the upper bound below. Proofs of the auxiliary results and the lower bound are given in the Supplemental Material (SM)~\footnote{See the Supplemental Material appended below for the proofs, the explicit coordinates and pullbacks of the examples, and the complexity analysis, which includes Ref.~\cite{FuchsVanDeGraaf1999}.}.

\textit{Classical labels.}---Let \(\{\cV_x\}_{x\in\Gamma}\) be unitary channels \(\cV_x(\rho)=V_x\rho V_x^\dagger\). The decoder
\begin{equation}
  \cC_\Gamma(Z):=\sum_{x\in\Gamma}V_x\Tr_P\bigl[(I\otimes\Ket x\!\Bra x)Z(I\otimes\Ket x\!\Bra x)\bigr]V_x^\dagger
  \label{eq:decoder}
\end{equation}
measures the program register in the label basis and applies \(V_x\). For a classical program, \(\cC_\Gamma(\,\cdot\,\otimes\sigma_u)=\sum_xq(x\mid u)\cV_x\), so the implemented channel is the barycenter of the label channels with weights \(q(\,\cdot\mid u)\). The approximation error is determined by the barycenter of the label channels. A deterministic choice, such as the nearest point of a net, generally gives an error linear in the mesh spacing. Unbiased rounding assigns probabilities to nearby labels so that their first-order deviations from the target cancel in the barycenter, leaving an error quadratic in the mesh spacing.

\textit{Unbiased rounding.}---We first consider coordinates \(F:B\to\cS\) given on a single rectangle \(B=\prod_{j=1}^\nu[a_j,b_j]\subset\R^\nu\), where \(F\) is surjective and \(C^2\) as a map into the Hermiticity-preserving maps. Let \(\Gamma_h\subset B\) be the product lattice that divides each side of \(B\) into cells of width at most \(h\) (Fig.~\ref{fig:lattice}), put \(\cV_x:=F(x)\) for \(x\in\Gamma_h\), and let \(t\in B\) be coordinates of the target, \(F(t)=\cU_u\). For each \(j\), let \(x_j^-\le t_j\le x_j^+\) be the adjacent lattice points and write \(t_j=(1-r_j)x_j^-+r_jx_j^+\). Independently for each \(j\), let \(X_j=x_j^+\) with probability \(r_j\) and \(X_j=x_j^-\) otherwise. The distribution \(q_h(\,\cdot\mid u)\) of \(X\) satisfies \(\E[X]=t\) and \(\E\norm{X-t}_2^2\le\nu h^2/4\). The Taylor expansion \(F(X)=F(t)+DF_t[X-t]+R\) has a remainder \(\norm{R}_\diamond\le\frac12M_2\norm{X-t}_2^2\), where \(M_2\) is a uniform bound on the second derivative of \(F\) in the diamond norm. The first-order term vanishes on average, and hence
\begin{equation}
  \tfrac12\Bigl\lVert\sum_{x\in\Gamma_h}q_h(x\mid u)\,\cV_x-\cU_u\Bigr\rVert_\diamond
  \le\frac{M_2\nu}{16}h^2 .
  \label{eq:rounding}
\end{equation}
The mesh \(h=4\sqrt{\eps/(M_2\nu)}\) gives error \(\eps\) with \(\abs{\Gamma_h}=O(h^{-\nu})=O(\eps^{-\nu/2})\) labels, that is, a program cost of \(\frac\nu2\log(1/\eps)+O(1)\). Deterministic rounding to the nearest lattice point would require \(h=O(\eps)\) and \(O(\eps^{-\nu})\) labels.

A general \(M\) need not admit coordinates on a single rectangle. Since \(M\) is compact, finitely many rectangles \(B_\alpha\) and maps \(F_\alpha\) cover \(\cS\) (Fig.~\ref{fig:lattice}). For each target \(u\in M\), we first select a rectangle \(B_\alpha\) whose image contains \(\cU_u\), and then choose \(t\in B_\alpha\) satisfying \(F_\alpha(t)=\cU_u\). If more than one rectangle or coordinate is available, we use a fixed rule to select one pair \(s(u)=(\alpha,t)\), as described in the proof of Lemma~\ref{lem:atlas} in the SM.
Unbiased rounding is then applied to \(t\) within the selected rectangle, yielding a probability distribution over its lattice points \(x\). We use the corresponding distribution over labels \((\alpha,x)\) as the classical program. Equation~\eqref{eq:rounding} holds with \(M_2\) chosen as a common bound on the second derivatives of all \(F_\alpha\). Such a finite bound exists because the second derivatives are continuous on compact boxes and there are only finitely many charts.
This proves the upper bound of Theorem~\ref{thm:main} for a known target.

\textit{Barycentric programming.}---More generally, for a probability distribution \(p(\dd v\mid u)\) on \(M\), consider the classical program with distribution
\begin{equation}
  q(x\mid u)=\int q_h(x\mid v)\,p(\dd v\mid u)
  \label{eq:barycentric}
\end{equation}
and the decoder \(\cC_{\Gamma_h}\). We call this pair the barycentric programming by \(p\). By Eq.~\eqref{eq:decoder}, it implements the channel \(\cE_u=\int\sum_xq_h(x\mid v)\cV_x\,p(\dd v\mid u)\). Writing \(\overline{\cU}_u:=\int\cU_v\,p(\dd v\mid u)\) for the averaged channel, the triangle inequality gives
\begin{equation}
  \begin{aligned}
    \tfrac12\norm{\cE_u-\cU_u}_\diamond
    &\le\tfrac12\norm{\cE_u-\overline{\cU}_u}_\diamond
    +\tfrac12\norm{\overline{\cU}_u-\cU_u}_\diamond\\
    &\le\frac{M_2\nu}{16}h^2
    +\tfrac12\norm{\overline{\cU}_u-\cU_u}_\diamond .
  \end{aligned}
  \label{eq:bary-error}
\end{equation}
The second inequality follows from Eq.~\eqref{eq:rounding} and convexity of the diamond norm. The remaining term depends only on the averaged channel \(\overline{\cU}_u\). The known case is \(p=\delta_u\). For an unknown target, \(p\) is the distribution of an estimate pulled back onto the family.

\textit{Unknown target.}---When \(\cU_u\) is given as a black box, the encoder proceeds in three stages,
\begin{equation}
  \begin{aligned}
    \cU_u
    &\xrightarrow{\ \text{estimation}\ }
    \widehat u\in\PU(d)
    \xrightarrow{\ \text{pullback}\ }
    \Pi(\widehat u)\in M\\
    &\xrightarrow{\ \text{rounding}\ }
    x\in\Gamma_h .
  \end{aligned}
  \label{eq:pipeline}
\end{equation}
The rounding stage draws and stores a label according to \(q_h(\,\cdot\mid\Pi(\widehat u))\), so Eq.~\eqref{eq:barycentric} applies with \(p\) equal to the distribution of \(\Pi(\widehat u)\). For the estimation we use the estimator of Haah et al.~\cite{HaahKothariODonnellTang2023}, which uses at most \(n\) adaptive queries, a single qudit, and computational-basis measurements. Its output distribution \(p_n(\dd\widehat u\mid u)\) satisfies
\begin{equation}
  \E\,d_F(u,\widehat u)^2\le\frac{C}{n^2},
  \qquad
  \norm{\E[\cU_{\widehat u}]-\cU_u}_\diamond\le\frac{C}{n^2},
  \label{eq:haah}
\end{equation}
with \(C\) depending only on \(d\). Here \(d_F(u,v):=(1-\abs{\Tr(U^\dagger V)}^2/d^2)^{1/2}\) is the trace distance between the Choi states~\cite{Choi1975} of \(\cU_u\) and \(\cU_v\), and it is equivalent to the diamond distance up to a factor \(\sqrt{2d}\) (see Lemma~\ref{lem:dF-diamond} in the SM). The first inequality expresses Heisenberg scaling, known for unitary estimation~\cite{Hayashi2006,Kahn2007}, and the second bounds the bias of the estimated channel by the same order as its mean squared error. A derivation of Eq.~\eqref{eq:haah} is given in the SM (Lemma~\ref{lem:haah}).

The estimate \(\widehat u\) generally lies outside \(M\). We pull it back to \(M\) using a retraction \(\Pi:\PU(d)\to M\), a map that is the identity on \(M\) and is \(C^2\) as a map of channels on a neighborhood of \(M\).
Such a local \(C^2\) retraction exists by the tubular neighborhood theorem~\cite{LeeRiemannian2018}. We extend it outside this neighborhood by mapping all remaining points to a fixed point of \(M\).

\begin{lemma}
\label{lem:pullback}
There are \(C<\infty\) and \(n_0\ge1\) depending only on \(M\) and \(\Pi\) such that, for all \(n\ge n_0\),
\begin{equation}
  \sup_{u\in M}\tfrac12\norm{\E[\cU_{\Pi(\widehat u)}]-\cU_u}_\diamond\le\frac{C}{n^2}.
  \label{eq:pullback}
\end{equation}
\end{lemma}

The full proof is given in the SM (Lemma~\ref{lem:retraction}).

For \(0<h\le1\) and \(n\ge n_0\), substituting Eq.~\eqref{eq:pullback} into Eq.~\eqref{eq:bary-error} bounds the error by \(M_2\nu h^2/16+C/n^2\). The pullback and rounding are specified independently of \(u\) and require no further queries. Thus the entire procedure is a \(u\)-independent strategy using at most \(n\) queries. For sufficiently small \(\eps\), choose \(h=\Theta(\sqrt\eps)\) and \(n=\Theta(\eps^{-1/2})\) so that each error term is at most \(\eps/2\). The resulting classical program has error at most \(\eps\) and uses at most \(\frac\nu2\log(1/\eps)+O(1)\) bits. This completes the proof of the upper bound of Theorem~\ref{thm:main}.

\textit{Examples.}---For \(M=\PU(d)\), the coordinates \(F(t)=\cU_{\exp(i\sum_at_aE_a)}\), with a Hilbert--Schmidt orthonormal basis \(\{E_a\}\) of the traceless Hermitian matrices, cover \(\PU(d)\) from the single rectangle \([-\pi\sqrt d,\pi\sqrt d]^{d^2-1}\), and no pullback is needed. When the unitary family consists of phase gates \(\Ket0\!\Bra0+e^{-i\theta}\Ket1\!\Bra1\), we use the following pullback. For estimates sufficiently close to this family, it sets the off-diagonal entries of the estimated unitary matrix \(\widehat U\) to zero and replaces each diagonal entry \(z\) by \(z/\abs z\). When the unitary family consists of reflections \(I-2\Ket\psi\!\Bra\psi\), each unitary has eigenvalue \(-1\) along \(\Ket\psi\) and eigenvalue \(+1\) on its orthogonal complement. After aligning the global phase of a sufficiently accurate estimate, the pullback returns \(I-2\lvert\widehat\psi\rangle\!\langle\widehat\psi\rvert\), where \(\lvert\widehat\psi\rangle\) is a normalized eigenvector of the estimate with eigenvalue near \(-1\). The coordinates and pullbacks of these families are given explicitly in the SM.

\textit{Efficiency.}---The decoder applies \(V_x\), which can be compiled into \(\operatorname{poly}(d)\operatorname{polylog}(1/\eps)\) gates by a two-level decomposition and the Solovay--Kitaev theorem~\cite{NielsenChuangBook,Kitaev1997,DawsonNielsen2006}. Given coordinates of the point to be rounded, the encoder needs \(O(\nu)\) arithmetic operations on \(\Theta(\log(1/\eps))\)-bit numbers and \(O(\nu\log(1/\eps))\) random bits, comparable to the label length. The rounding tolerates errors of \(O(\eps)\) in the coordinates, measured by the diamond distance of the corresponding channels, and errors of \(O(\sqrt\eps)\) in the rounding probabilities. The complexity and precision bounds are detailed in Sec.~\ref{sec:complexity} of the SM.

\textit{Lower bound.}---The lower bound follows the argument of YRC~\cite{YangRennerChiribella2020}, with the representation-theoretic counting replaced by a communication bound. (i) For any processor with error \(\eps\), the recycling construction of YRC provides a decoder that uses a single copy of the program state to implement \(\cU_u^{\otimes m}\) with error \(O(m\sqrt\eps)\)~\cite{YangRennerChiribella2020,GschwendtnerWinter2021}. (ii) From \(m\) parallel uses of \(\cU_u\), the \(\nu\) local coordinates of \(u\) can be estimated with mean squared error \(O(m^{-2})\), by the parallel estimator of YRC and the equivalence between the coordinate distance and \(d_F\) on a compact chart. (iii) The program register therefore transmits \(\nu\) real parameters with precision \(1/m\), and the communication bound of Yang, Chiribella, and Hayashi~\cite{YangChiribellaHayashi2020} requires \(\dim\cH_P\gtrsim m^\nu\). Keeping the error \(O(m\sqrt\eps)\) of step (i) small requires \(m\) to be smaller than \(\eps^{-1/2}\) by a factor of order \(\log(1/\eps)\), which produces the correction \(\frac{3\nu}2\log\log(1/\eps)\) in Eq.~\eqref{eq:main}. None of these steps uses group structure, so the bound holds for every family satisfying our assumption.
The full proof is given in Sec.~\ref{sec:lower} of the SM.

\textit{Conclusion.}---We have shown that a finite classical program can attain the optimal leading-order memory cost of programming a quantum processor. Barycentric programming achieves this by randomly selecting a label so that first-order approximation errors cancel in the implemented channel, halving the leading memory cost relative to deterministic rounding. For the smooth \(\nu\)-parameter families considered here, the resulting cost is \((\nu/2)\log_2(1/\eps)+O(1)\) classical bits. Together with our lower bound for arbitrary quantum programs, this establishes the leading-order scaling conjectured by YRC~\cite{YangRennerChiribella2020}. The construction also applies to unitary families that do not form a group. Even when the target unitary is unknown, the program can be generated using \(O(\eps^{-1/2})\) queries to it. Once generated, the program can be stored in classical memory and transmitted over a classical channel, with no quantum memory needed between its generation and use.

\begin{acknowledgments}
This work was supported by the MEXT Quantum Leap Flagship Program (MEXT QLEAP) JPMXS0118069605, Japan Society for the Promotion of Science (JSPS) KAKENHI Grant No.~26K25550, JST CREST Grant No.~JPMJCR25I5, JST ASPIRE JPMJAP25A3, JST NEXUS JPMJNX26C9, and IBM Quantum. We used ChatGPT~5.6 Pro (OpenAI) to derive the proof of Lemma~\ref{lem:pullback}. We also used Claude Code (Anthropic) with the models Claude Opus~5, Claude Opus~5.5, Claude Sonnet~5.5, and Claude Fable~5 to prepare the manuscript, including drafting, revision, and translation of the text, to draw the TikZ figures in the main text and the Supplemental Material, and to search for references and check their bibliographic data. The authors directed these tools with written instructions on the content and structure of each revision and verified all mathematical statements, references, and text.
\end{acknowledgments}

\bibliography{classical_compression}

\clearpage
\onecolumngrid
\setcounter{section}{0}
\setcounter{equation}{0}
\setcounter{figure}{0}
\setcounter{definition}{0}
\setcounter{secnumdepth}{2}
\renewcommand{\thesection}{S\arabic{section}}
\renewcommand{\theequation}{S\arabic{equation}}
\renewcommand{\thefigure}{S\arabic{figure}}
\renewcommand{\thedefinition}{S\arabic{definition}}
\renewcommand{\theHequation}{S\arabic{equation}}
\renewcommand{\theHfigure}{S\arabic{figure}}
\begin{center}
\large\bfseries
Appendix
\end{center}

Section~\ref{sec:prelim} fixes the notation. Sections~\ref{sec:atlas}--\ref{sec:unknown} prove the upper bound of Theorem~\ref{thm:main}, Sec.~\ref{sec:examples} treats the examples, Sec.~\ref{sec:complexity} analyzes the computational complexity, and Sec.~\ref{sec:lower} proves the lower bound.

\section{Preliminaries}
\label{sec:prelim}

All Hilbert spaces are finite dimensional, \(\cH\simeq\C^d\), and logarithms are to base 2. We write \(\HP(\cH)\) for the real vector space of Hermiticity-preserving maps on \(\cB(\cH)\) and \(\norm{\,\cdot\,}_\diamond\) for the diamond norm. For \(u=[U]\in\PU(d)\) we write \(\cU_u(\rho)=U\rho U^\dagger\). For a map \(G\) on an open set \(\Omega\) of a finite-dimensional real normed space, \(DG_t\) and \(D^2G_t\) denote the first and second Fr\'echet derivatives at \(t\); \(DG_t[v]\) is the directional derivative in the direction \(v\), and \(D^2G_t[v,w]\) is the second directional derivative in the directions \(v,w\).

\paragraph*{Taylor formula.}
Let \(G\) be \(C^2\), and suppose that the segment joining \(t,x\in\Omega\) lies in \(\Omega\). Applying the fundamental theorem of calculus twice to \(g(r)=G(t+r(x-t))\) gives
\begin{equation}
  G(x)=G(t)+DG_t[x-t]
  +\int_0^1(1-r)\,D^2G_{t+r(x-t)}[x-t,x-t]\,\dd r .
  \label{eq:taylor}
\end{equation}
Since \(\int_0^1(1-r)\,\dd r=\frac12\), a bound \(M_2\) on the second derivative bounds the remainder by \(\frac12M_2\norm{x-t}^2\).

\paragraph*{The distance \(d_F\).}
We write the maximally entangled state, the Choi vector of \(U\), and the Choi state~\cite{Choi1975} of \(u=[U]\) as
\[
  \Ket\Phi:=\frac1{\sqrt d}\sum_{i=1}^d\Ket i\Ket i,
  \qquad
  \Ket{\Phi_U}:=(U\otimes I)\Ket\Phi,
  \qquad
  J(u):=\Ket{\Phi_U}\!\Bra{\Phi_U}.
\]
The distance
\begin{equation}
  d_F(u,v)
  :=\sqrt{1-\frac{\abs{\Tr(U^\dagger V)}^2}{d^2}},
  \qquad u=[U],\quad v=[V],
  \label{eq:dF-def}
\end{equation}
satisfies, by \(\abs{\langle\Phi_U\vert\Phi_V\rangle}=\abs{\Tr(U^\dagger V)}/d\) and the formula for the trace distance between pure states~\cite{FuchsVanDeGraaf1999},
\begin{equation}
  d_F(u,v)
  =\frac12\norm{J(u)-J(v)}_1
  =\frac1{\sqrt2}\norm{J(u)-J(v)}_2 ,
  \label{eq:dF-choi}
\end{equation}
where \(\norm{\,\cdot\,}_2\) is the Hilbert--Schmidt norm. In particular, \(d_F\) does not depend on the phases of the representatives and is bi-invariant.

\begin{smlemma}
\label{lem:dF-diamond}
For all \(u,v\in\PU(d)\),
\begin{equation}
  d_F(u,v)
  \le\frac12\norm{\cU_u-\cU_v}_\diamond
  \le\sqrt{2d}\,d_F(u,v).
  \label{eq:dF-diamond}
\end{equation}
\end{smlemma}

\begin{proof}
For the left inequality, choose \(\Ket\Phi\) as the input in the diamond norm and use Eq.~\eqref{eq:dF-choi}. For unitaries \(A,B\) and a state \(\rho\), \(A\rho A^\dagger-B\rho B^\dagger=(A-B)\rho A^\dagger+B\rho(A-B)^\dagger\), so \(\norm{A\rho A^\dagger-B\rho B^\dagger}_1\le2\norm{A-B}_\infty\). Putting \(A=U\otimes I\) and \(B=e^{i\phi}V\otimes I\), we obtain
\[
  \frac12\norm{\cU_u-\cU_v}_\diamond
  \le\norm{U-e^{i\phi}V}_\infty
  \le\norm{U-e^{i\phi}V}_2 .
\]
Choosing \(\phi\) with \(e^{i\phi}\Tr(U^\dagger V)=\abs{\Tr(U^\dagger V)}\) and putting \(x:=\abs{\Tr(U^\dagger V)}/d\in[0,1]\), we have \(\norm{U-e^{i\phi}V}_2^2=2d(1-x)\le2d(1-x^2)=2d\,d_F(u,v)^2\).
\end{proof}

\section{Finite rectangular parametrizations}
\label{sec:atlas}

Throughout, \(M\subset\PU(d)\) satisfies the assumption of the main text:
\begin{itemize}
\item[(R1)] \(M\) is a compact boundaryless \(C^3\) submanifold of \(\PU(d)\) of dimension \(\nu\ge1\),
\end{itemize}
and \(\cS=\{\cU_u\}_{u\in M}\). We cover \(\cS\) by smooth maps from finitely many rectangles.

\begin{definition}[Finite rectangular parametrization]
\label{def:rect}
A \(\nu\)-dimensional finite rectangular parametrization of \(\cS\) consists of the following data.
\begin{itemize}
\item[(R2)] Rectangles \(B_\alpha=\prod_{j=1}^\nu[a_{\alpha,j},b_{\alpha,j}]\subset\R^\nu\) (\(\alpha=1,\ldots,L\)) and \(C^2\) maps \(F_\alpha\) into \(\HP(\cH)\), defined on open sets slightly enlarging \(B_\alpha\), such that \(F_\alpha(B_\alpha)\subset\cS\) and
\[
  \cS=\bigcup_{\alpha=1}^LF_\alpha(B_\alpha);
\]
together with a Borel measurable map \(s(u)=(\alpha,t)\), called a selector, that chooses for each \(u\in M\) one coordinate with \(t\in B_\alpha\) and \(F_\alpha(t)=\cU_u\).
\end{itemize}
\end{definition}

The maps \(F_\alpha\) need not be injective, and several coordinates may represent the same channel on the boundaries of the rectangles or at coordinate singularities. The selector need not be continuous. When a single smooth surjection \(F:B\to\cS\) is available (\(L=1\)), we call \(B\) a global parameter box.

\begin{smlemma}
\label{lem:atlas}
If \(M\) satisfies (R1), then \(\cS\) has a \(\nu\)-dimensional finite rectangular parametrization.
\end{smlemma}

\begin{proof}
The map \(u\mapsto\cU_u\) is a smooth embedding of \(\PU(d)\) into \(\HP(\cH)\). At each point \(u\in M\), take a \(C^3\) coordinate chart \(\kappa\) of \(M\), and choose in the coordinate image a compact rectangle that contains a neighborhood of \(\kappa(u)\) and lies in the interior of the coordinate neighborhood. The preimages of the interiors of these rectangles cover \(M\), so by compactness finitely many rectangles \(B_1,\ldots,B_L\) can be chosen. The maps \(F_\alpha:=\cU\circ\kappa_\alpha^{-1}\) satisfy the first half of (R2). As the selector, choose the smallest \(\alpha\) with \(u\in\kappa_\alpha^{-1}(B_\alpha)\) and put \(t=\kappa_\alpha(u)\). This is a case distinction by finitely many compact sets and is therefore Borel.
\end{proof}

\section{Known target}
\label{sec:known}

\subsection{Unbiased lattice rounding}

Fix a finite rectangular parametrization and a mesh \(0<h\le1\). For each coordinate of each \(B_\alpha\) let
\[
  N_{\alpha,j}=\ceil{(b_{\alpha,j}-a_{\alpha,j})/h},
  \qquad
  \delta_{\alpha,j}=(b_{\alpha,j}-a_{\alpha,j})/N_{\alpha,j}\le h,
\]
and let \(\Gamma_{\alpha,h}\subset B_\alpha\) be the product lattice with vertices \(a_{\alpha,j}+k\delta_{\alpha,j}\) (\(0\le k\le N_{\alpha,j}\)). The label set and the channels of the labels are
\[
  \Gamma_h:=\bigsqcup_{\alpha=1}^L\{\alpha\}\times\Gamma_{\alpha,h},
  \qquad
  \cV_{(\alpha,x)}:=F_\alpha(x),
\]
and each \(\cV_{(\alpha,x)}\) is a unitary channel since \(F_\alpha(x)\in\cS\).

For \(u\in M\) let \(s(u)=(\alpha,t)\). For each coordinate \(j\) let \(x_j^-\le t_j\le x_j^+\) be the adjacent lattice points enclosing \(t_j\), and write \(t_j=(1-r_j)x_j^-+r_jx_j^+\). Independently for each \(j\), let \(X_j=x_j^-\) with probability \(1-r_j\) and \(X_j=x_j^+\) with probability \(r_j\), and let \(q_h(\,\cdot\mid u)\) be the distribution of the label \((\alpha,X)\).

\begin{smlemma}
\label{lem:rounding}
\(\E[X]=t\) and \(\E\norm{X-t}_2^2\le\nu h^2/4\).
\end{smlemma}

\begin{proof}
We have \(\E[X_j]=(1-r_j)x_j^-+r_jx_j^+=t_j\) and \(\E[(X_j-t_j)^2]=r_j(1-r_j)\delta_{\alpha,j}^2\le h^2/4\). Summing over \(j\) gives the second claim.
\end{proof}

\begin{smlemma}
\label{lem:taylor}
Let \(B\subset\R^\nu\) be a rectangle, and let \(F\) be a \(C^2\) map satisfying \(\norm{D^2F_s[v,w]}_\diamond\le M_2\norm v_2\norm w_2\) on \(B\). If a \(B\)-valued random variable \(X\) has finite support and \(\E[X]=t\), then
\[
  \norm{\E[F(X)]-F(t)}_\diamond
  \le\tfrac12 M_2\,\E\norm{X-t}_2^2 .
\]
\end{smlemma}

\begin{proof}
Since the segment joining \(X\) and \(t\) lies in \(B\), Eq.~\eqref{eq:taylor} gives \(F(X)=F(t)+DF_t[X-t]+R\) with \(\norm{R}_\diamond\le\frac12M_2\norm{X-t}_2^2\). Taking expectations, the first-order term vanishes because \(DF_t\) is linear and \(\E[X-t]=0\).
\end{proof}

We define the uniform bound on the second derivatives by
\begin{equation}
  M_2
  :=\max_{\alpha}\sup_{t\in B_\alpha}
    \sup_{\norm{v}_2=\norm{w}_2=1}
    \norm{D^2F_{\alpha,t}[v,w]}_\diamond ,
  \label{eq:M2}
\end{equation}
which is finite since each \(B_\alpha\) is compact.

\begin{proposition}
\label{prop:rounding}
For \(0<h\le1\), the label set \(\Gamma_h\) and the distribution \(q_h\) satisfy
\[
  \tfrac12\Bigl\lVert\sum_{x\in\Gamma_h}q_h(x\mid u)\,\cV_x-\cU_u\Bigr\rVert_\diamond
  \le\frac{M_2\nu}{16}h^2
  \qquad(u\in M),
  \qquad
  \abs{\Gamma_h}\le C_\Gamma h^{-\nu},
\]
where \(C_\Gamma:=\sum_\alpha\prod_j(b_{\alpha,j}-a_{\alpha,j}+2)\).
\end{proposition}

\begin{proof}
If \(s(u)=(\alpha,t)\), then \(\sum_xq_h(x\mid u)\cV_x=\E[F_\alpha(X)]\) and \(\cU_u=F_\alpha(t)\). By Lemmas~\ref{lem:rounding} and~\ref{lem:taylor}, the error is at most \(\frac12\cdot\frac{M_2}{2}\cdot\frac{\nu h^2}{4}\). The bound on the number of labels follows from \(N_{\alpha,j}+1\le(b_{\alpha,j}-a_{\alpha,j})/h+2\le(b_{\alpha,j}-a_{\alpha,j}+2)/h\).
\end{proof}

\subsection{Programs of classical labels}

For each \(x\in\Gamma_h\) choose a unitary \(V_x\) with \(\cV_x(\rho)=V_x\rho V_x^\dagger\). For each \(u\in M\) fix a probability distribution \(q(\,\cdot\mid u)\) on \(\Gamma_h\), and define the program on \(\cH_P:=\operatorname{span}\{\Ket x\}_{x\in\Gamma_h}\) and the decoder \(\cC_{\Gamma_h}\) of Eq.~\eqref{eq:decoder} by
\[
  \sigma_u:=\sum_{x\in\Gamma_h}q(x\mid u)\,\Ket x\!\Bra x,
  \qquad
  \cC_{\Gamma_h}(Z)
  :=\sum_{x\in\Gamma_h}
    V_x\,
    \Tr_P\bigl[(I\otimes\Ket x\!\Bra x)Z(I\otimes\Ket x\!\Bra x)\bigr]
    V_x^\dagger .
\]

\begin{smlemma}
\label{lem:label}
The map \(\cC_{\Gamma_h}\) is CPTP, and \(\cC_{\Gamma_h}(\,\cdot\,\otimes\sigma_u)=\sum_{x\in\Gamma_h}q(x\mid u)\,\cV_x\).
\end{smlemma}

\begin{proof}
Each term is completely positive, and the trace is preserved because the projectors \(\Ket x\!\Bra x\) sum to the identity. Since \(\sigma_u\) is diagonal, \(\Tr_P[(I\otimes\Ket x\!\Bra x)(\rho\otimes\sigma_u)(I\otimes\Ket x\!\Bra x)]=q(x\mid u)\rho\).
\end{proof}

\begin{smtheorem}
\label{thm:known}
Suppose that \(\cS\) has a \(\nu\)-dimensional finite rectangular parametrization with \(M_2>0\). For \(\eps\le M_2\nu/16\), the program with distribution \(q_h(\,\cdot\mid u)\) and the decoder \(\cC_{\Gamma_h}\) form an \(\eps\)-processor, and
\[
  c_P^*(\eps;\cS)
  \le\frac\nu2\log\frac1\eps
  +\frac\nu2\log\frac{M_2\nu}{16}
  +\log C_\Gamma .
\]
If \(M_2=0\), the family can be implemented exactly with a fixed finite register.
\end{smtheorem}

\begin{proof}
Put \(h=4\sqrt{\eps/(M_2\nu)}\le1\). By Lemma~\ref{lem:label} and Proposition~\ref{prop:rounding}, the error is at most \(\eps\), and \(\log\abs{\Gamma_h}\le\log C_\Gamma+\nu\log(1/h)\). If \(M_2=0\), the error in Proposition~\ref{prop:rounding} vanishes for every \(h\).
\end{proof}

By Lemma~\ref{lem:atlas}, Theorem~\ref{thm:known} proves the upper bound of Eq.~\eqref{eq:main} with a classical program over \(O(\eps^{-\nu/2})\) labels.

\section{Unknown target}
\label{sec:unknown}

We now prove the estimation and pullback bounds used to generate the program from queries to \(\cU_u\). The label set \(\Gamma_h\) and the decoder \(\cC_{\Gamma_h}\) are those of Sec.~\ref{sec:known}, and only the distribution of the label changes, according to the three stages of Eq.~\eqref{eq:pipeline}.

\subsection{Estimation}

\begin{smlemma}[Estimator of Haah et al.~\cite{HaahKothariODonnellTang2023}]
\label{lem:haah}
There are \(C<\infty\) and \(n_d\ge1\) depending only on \(d\) such that for every \(n\ge n_d\), a \(u\)-independent adaptive strategy using at most \(n\) queries to \(\cU_u\) outputs an estimate \(\widehat u\in\PU(d)\) whose distribution \(p_n(\dd\widehat u\mid u)\) satisfies
\begin{align}
  \int d_F(u,\widehat u)^2\,p_n(\dd\widehat u\mid u)
  &\le\frac{C}{n^2},
  \label{eq:haah-mse}\\
  \left\lVert
    \int\cU_{\widehat u}\,p_n(\dd\widehat u\mid u)-\cU_u
  \right\rVert_\diamond
  &\le\frac{C}{n^2}.
  \label{eq:haah-bias}
\end{align}
The estimation can be carried out with a single qudit, computational-basis measurements, and \(\operatorname{poly}(d,n)\) known gates and classical operations.
\end{smlemma}

\begin{proof}
Apply Theorem~1.1 of Ref.~\cite{HaahKothariODonnellTang2023} with accuracy \(\delta=\kappa_d/n\), where \(\kappa_d\) is chosen so that the query bound \(O(d^2/\delta)\) is at most \(n\). The fourth guarantee of that theorem (the mean squared error in diamond distance) together with Eq.~\eqref{eq:dF-diamond} gives Eq.~\eqref{eq:haah-mse}, and the fifth guarantee gives Eq.~\eqref{eq:haah-bias}. The claims on the implementation follow from the second and third guarantees.
\end{proof}

\subsection{Retractions}

The family of channels \(\cS\) is a compact \(C^3\) submanifold of \(\HP(\cH)\). We write \(\operatorname{dist}_\diamond(X,\cS):=\min_{Y\in\cS}\norm{X-Y}_\diamond\) for \(X\in\HP(\cH)\). Fixing an inner product on \(\HP(\cH)\), the tubular neighborhood theorem shows that the nearest-point projection with respect to that inner product is uniquely defined and \(C^2\) on some open neighborhood of \(\cS\)~\cite{LeeRiemannian2018}. Since \(\cS\) is compact, for some \(\rho>0\) the set
\[
  N_\rho(\cS):=\Set{X\in\HP(\cH):\operatorname{dist}_\diamond(X,\cS)<\rho}
\]
is contained in this open neighborhood. More generally, consider a \(C^2\) map \(\pi:N_\rho(\cS)\to\cS\) with \(\pi|_\cS=\id\). Fix a point \(u_\ast\in M\) and define
\begin{equation}
  \Pi(\widehat u):=
  \begin{cases}
    \text{the unique \(v\in M\) with \(\cU_v=\pi(\cU_{\widehat u})\)},
      & \cU_{\widehat u}\in N_\rho(\cS),\\
    u_\ast,
      & \text{otherwise}.
  \end{cases}
  \label{eq:retraction}
\end{equation}
We call \(\Pi\) a retraction. Since \(v\mapsto\cU_v\) is an embedding, the first branch is continuous on an open set, and \(\Pi\) is Borel measurable. When a \(C^2\) map \(\Pi_0\) that is defined on a neighborhood of \(M\) in \(\PU(d)\) and is the identity on \(M\) is given, composing it with the nearest-point projection onto the set of all unitary channels \(\{\cU_w\}_{w\in\PU(d)}\) also yields a map \(\pi\) satisfying the above conditions. The pullbacks of Sec.~\ref{sec:examples} are of this form.

\begin{smlemma}[Lemma~\ref{lem:pullback}]
\label{lem:retraction}
Fix a retraction \(\Pi\). There are \(C<\infty\) and \(n_0\ge1\) depending only on \(M\) and \(\Pi\) such that for all \(n\ge n_0\),
\[
  \sup_{u\in M}
  \tfrac12\Bigl\lVert
    \int\cU_{\Pi(\widehat u)}\,p_n(\dd\widehat u\mid u)
    -\cU_u
  \Bigr\rVert_\diamond
  \le\frac{C}{n^2}.
\]
\end{smlemma}

\begin{figure}[t]
\centering
\begin{tikzpicture}[
  x=1cm, y=1cm,
  every node/.style={font=\small},
  pt/.style={circle,fill,inner sep=1.5pt},
  manifold/.style={draw=blue!60!black,line width=1.1pt},
  tangent/.style={draw=orange!85!black,line width=0.8pt,
    dash pattern=on 5pt off 2.5pt},
  disp/.style={-{Stealth[length=2.4mm]},gray!40!black,line width=0.9pt},
  tangential/.style={-{Stealth[length=2.6mm]},teal!55!black,line width=1.4pt},
  remainder/.style={-{Stealth[length=2.4mm]},violet!75!black,line width=1.4pt,
    dash pattern=on 1.2pt off 1.6pt},
  retraction/.style={-{Stealth[length=2.4mm]},red!70!black,line width=1.0pt,
    dash pattern=on 3pt off 2pt},
  helper/.style={gray!55,line width=0.5pt,dash pattern=on 1pt off 1.5pt},
  sample/.style={-{Stealth[length=1.5mm]},red!55!black,line width=0.45pt},
  estpt/.style={circle,draw=gray!45!black,fill=white,inner sep=1.3pt,line width=0.5pt},
  prjpt/.style={circle,fill=red!70!black,inner sep=1.1pt},
  leader/.style={gray!60,line width=0.45pt},
  panel/.style={rounded corners=5pt,draw=gray!45,fill=gray!2,line width=0.7pt},
  ttl/.style={font=\small\bfseries,anchor=north west},
  note/.style={font=\scriptsize,align=left,text=gray!20!black}
]

\path[panel] (0.00,0.00) rectangle (8.55,6.40);
\node[ttl] at (0.22,6.28) {(a) Decomposition for a single estimate};

\draw[tangent] (0.40,2.55) -- (7.95,4.815);
\draw[manifold] plot[domain=0.40:7.95,samples=80,smooth]
  (\x,{3.0+0.30*(\x-1.9)-0.07*(\x-1.9)^2});
\node[anchor=south east,text=orange!85!black] at (7.95,4.83) {$T_{\cU_u}\cS$};
\node[anchor=north east,text=blue!60!black] at (7.95,2.25) {$\cS$};

\coordinate (u)   at (1.900,3.000);
\coordinate (est) at (5.940,5.328);
\coordinate (lin) at (6.247,4.304);
\coordinate (prj) at (5.500,3.173);

\draw[helper] (est) -- (lin);

\draw[disp] (u) -- (est)
  node[pos=0.48,above left=0pt,text=gray!30!black] {$\Delta_u$};
\draw[tangential] (u) -- (lin)
  node[pos=0.60,above=0pt,sloped,text=teal!55!black,font=\scriptsize]
  {$D\pi_{\cU_u}[\Delta_u]$};
\draw[retraction] (est) -- (prj)
  node[pos=0.30,left=1pt,text=red!70!black] {$\pi$};
\draw[remainder] (lin) -- (prj)
  node[pos=0.45,right=2pt,text=violet!75!black] {$R_u$};

\node[pt] at (u) {};
\node[anchor=north west,inner sep=3pt] at (u) {$\cU_u$};
\node[pt,fill=gray!40!black] at (est) {};
\node[anchor=south,inner sep=3pt] at (est) {$\cU_{\widehat u}$};
\node[pt,fill=teal!55!black] at (lin) {};
\node[pt,fill=red!70!black] at (prj) {};
\node[anchor=north east,inner sep=2.5pt,text=red!70!black] at (prj)
  {$\cU_{\Pi(\widehat u)}$};

\node[note,anchor=north west,text width=8.05cm] at (0.25,1.85)
  {\(\cU_{\Pi(\widehat u)}-\cU_u
     =\textcolor{teal!55!black}{D\pi_{\cU_u}[\Delta_u]}
     +\textcolor{violet!75!black}{R_u}\)\\[3pt]
   \textcolor{teal!55!black}{\textbf{First-order term}}: the tangential component of \(\Delta_u\), linear in \(\Delta_u\).\\[1pt]
   \textcolor{violet!75!black}{\textbf{Remainder}}:
   \(\norm{R_u}_\diamond\le\frac K2\norm{\Delta_u}_\diamond^2\).};

\path[panel] (8.80,0.00) rectangle (15.25,6.40);
\node[ttl] at (9.02,6.28) {(b) Averaging over estimates};

\coordinate (c) at (12.00,3.95);
\draw[tangent,draw=orange!85!black!45] (10.50,3.68) -- (13.60,4.24);
\draw[manifold] plot[domain=10.60:13.45,samples=80,smooth]
  (\x,{3.95+0.18*(\x-12)-0.40*(\x-12)^2});
\node[anchor=east,text=blue!60!black] at (10.55,2.95) {$\cS$};
\draw[gray!55,dashed,line width=0.6pt] (c) circle[radius=1.2];

\foreach \ex/\ey/\px/\py in {
    13.220/4.211/12.955/3.757, 12.397/4.989/12.320/3.967,
    11.570/5.161/11.895/3.927, 11.086/4.552/11.523/3.773,
    10.911/3.592/11.062/3.429, 11.573/3.005/11.143/3.502,
    12.362/2.795/12.896/3.790, 12.881/3.295/13.132/3.641}{
  \draw[sample,shorten >=1.2pt,shorten <=1.4pt] (\ex,\ey) -- (\px,\py);
  \node[estpt] at (\ex,\ey) {};
  \node[prjpt] at (\px,\py) {};
}
\node[pt] at (c) {};
\node[star,star points=5,star point ratio=2.3,fill=black,inner sep=1.1pt]
  at (12.116,3.723) {};

\draw[leader] (12.06,4.01) -- (13.50,5.33);
\node[anchor=west,font=\scriptsize,inner sep=1.5pt] at (13.45,5.38) {$\cU_u$};
\draw[leader] (12.18,3.67) -- (13.35,2.66);
\node[anchor=west,font=\scriptsize,inner sep=1.5pt] at (13.30,2.62)
  {$\E_u[\cU_{\Pi(\widehat u)}]$};
\node[note,anchor=south west,text=gray!45!black] at (9.05,5.22) {radius $O(1/n)$};
\draw[leader] (10.45,5.28) -- (11.15,4.80);

\node[estpt] at (9.20,2.15) {};
\node[note,anchor=west] at (9.32,2.15) {estimate \(\cU_{\widehat u}\)};
\node[prjpt] at (11.55,2.15) {};
\node[note,anchor=west] at (11.67,2.15) {pullback \(\cU_{\Pi(\widehat u)}\)};

\node[note,anchor=north west,text width=6.05cm] at (9.02,1.72)
  {Each displacement is \(O(1/n)\), but on average\\[1pt]
   \textcolor{teal!55!black}{\textbf{first order}}:
   \(\norm{\E_u[\Delta_u]}_\diamond\le C/n^2\) (Eq.~\eqref{eq:haah-bias})\\[1pt]
   \textcolor{violet!75!black}{\textbf{remainder}}:
   \(\E_u\norm{R_u}_\diamond\le C/n^2\) (Eq.~\eqref{eq:haah-mse})\\[2pt]
   Hence \(\norm{\E_u[\cU_{\Pi(\widehat u)}]-\cU_u}_\diamond=O(n^{-2})\).};

\end{tikzpicture}
\caption{Geometry of the proof of Lemma~\ref{lem:retraction} (schematic two-dimensional cross section). (a) For a single estimate, the displacement \(\cU_{\Pi(\widehat u)}-\cU_u\) caused by the pullback is split into the tangential component of \(\Delta_u\) (the first-order term) and the remainder. (b) The estimates (open circles) spread by \(O(1/n)\) around \(\cU_u\), and each is pulled back to a point on \(\cS\) (red dots). The mean of the pulled-back points (star) deviates from \(\cU_u\) by \(O(n^{-2})\).}
\label{fig:retraction}
\end{figure}
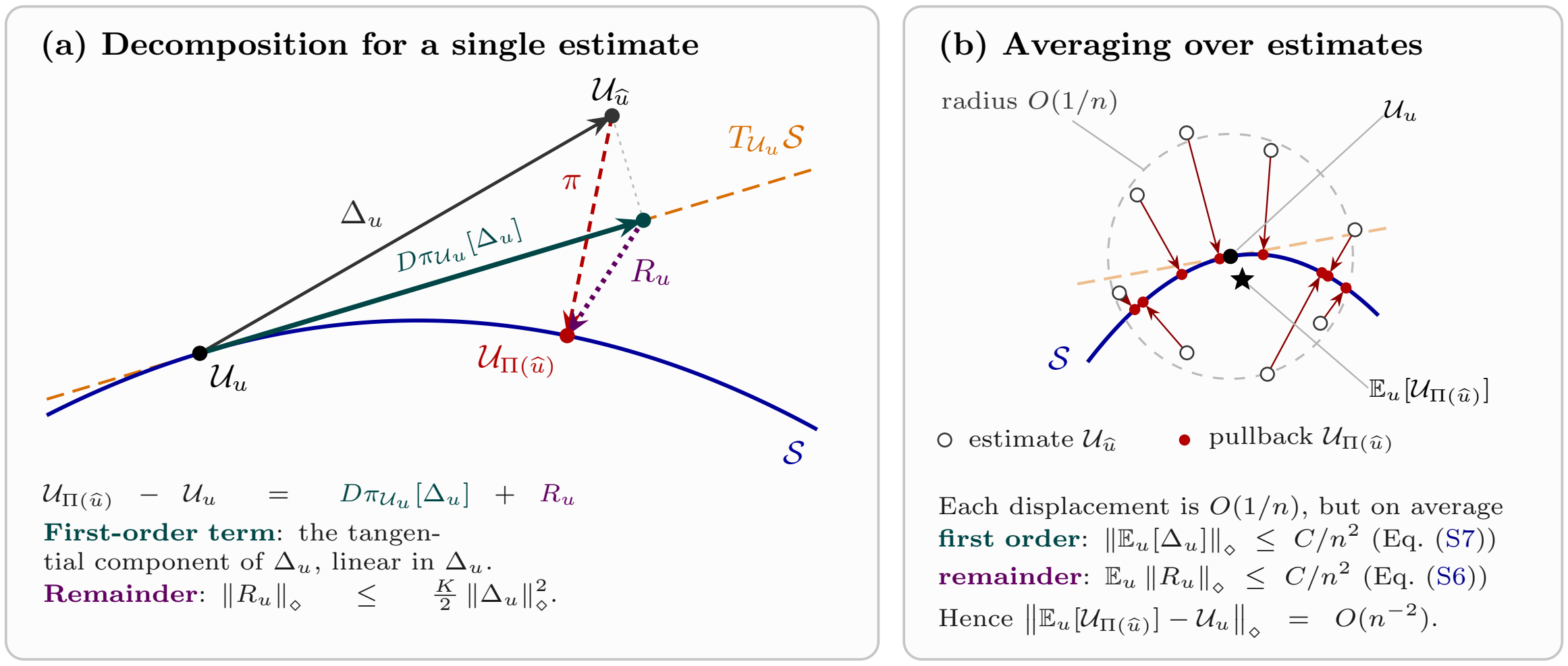

\begin{proof}
Fix \(n\ge n_d\) and \(u\in M\), and write \(\E_u\) and \(\Pr_u\) for the expectation and probability with respect to \(p_n(\dd\widehat u\mid u)\). Let \(\Delta_u(\widehat u):=\cU_{\widehat u}-\cU_u\) be the displacement of the channel caused by the estimation. Below, \(C\) denotes a constant that may change from line to line and depends only on \(d\), \(M\), and \(\pi\).

\textit{Two guarantees of the estimator.}
By Eq.~\eqref{eq:dF-diamond}, \(\norm{\Delta_u(\widehat u)}_\diamond\le2\sqrt{2d}\,d_F(u,\widehat u)\). Hence Eqs.~\eqref{eq:haah-mse} and~\eqref{eq:haah-bias} give
\begin{equation}
  \E_u\norm{\Delta_u}_\diamond^2\le\frac{C}{n^2},
  \qquad
  \norm{\E_u[\Delta_u]}_\diamond\le\frac{C}{n^2}.
  \label{eq:app-two}
\end{equation}

\textit{Derivatives of the retraction.}
Fixing \(0<r<\rho\), compactness gives
\[
  K:=\sup_{\operatorname{dist}_\diamond(X,\cS)\le r}
  \sup_{\norm A_\diamond,\norm B_\diamond\le1}
  \norm{D^2\pi_X[A,B]}_\diamond<\infty,
  \qquad
  K_1:=\sup_{u\in M}\norm{D\pi_{\cU_u}}_{\diamond\to\diamond}<\infty .
\]

\textit{Local Taylor expansion.}
Put \(B_r(u):=\{\widehat u:\norm{\Delta_u(\widehat u)}_\diamond<r\}\). If \(\widehat u\in B_r(u)\), the segment \(\cU_u+\tau\Delta_u\) (\(0\le\tau\le1\)) lies in \(N_\rho(\cS)\), so Eq.~\eqref{eq:taylor} and \(\pi(\cU_u)=\cU_u\) give
\begin{equation}
  \cU_{\Pi(\widehat u)}-\cU_u
  =D\pi_{\cU_u}[\Delta_u]+R_u,
  \qquad
  \norm{R_u}_\diamond\le\frac K2\norm{\Delta_u}_\diamond^2 .
  \label{eq:app-taylor}
\end{equation}
This decomposition is illustrated in Fig.~\ref{fig:retraction}.

\textit{Outside the neighborhood.}
Markov's inequality and Eq.~\eqref{eq:app-two} give \(\Pr_u(B_r(u)^c)\le C/(r^2n^2)\). Since \(\norm{\cU_v-\cU_w}_\diamond\le2\) for all \(v,w\in\PU(d)\),
\begin{equation}
  \norm{\E_u[\Delta_u\boldsymbol 1_{B_r(u)^c}]}_\diamond\le\frac{C}{n^2},
  \qquad
  \norm{\E_u[(\cU_{\Pi(\widehat u)}-\cU_u)\boldsymbol 1_{B_r(u)^c}]}_\diamond\le\frac{C}{n^2}.
  \label{eq:app-outside}
\end{equation}

\textit{Inside the neighborhood.}
The second inequality of Eq.~\eqref{eq:app-two} and the first of Eq.~\eqref{eq:app-outside} give \(\norm{\E_u[\Delta_u\boldsymbol 1_{B_r(u)}]}_\diamond\le C/n^2\). Averaging Eq.~\eqref{eq:app-taylor} over \(B_r(u)\) gives
\[
  \norm{\E_u[(\cU_{\Pi(\widehat u)}-\cU_u)\boldsymbol 1_{B_r(u)}]}_\diamond
  \le K_1\norm{\E_u[\Delta_u\boldsymbol 1_{B_r(u)}]}_\diamond
  +\frac K2\E_u\norm{\Delta_u}_\diamond^2
  \le\frac{C}{n^2}.
\]
Combining the two bounds, \(\norm{\E_u[\cU_{\Pi(\widehat u)}]-\cU_u}_\diamond\le C/n^2\) holds uniformly in \(u\).
\end{proof}

\section{Examples}
\label{sec:examples}

For \(\PU(d)\), the qubit phase gates, and the reflections, we give a global parameter box \((B,F)\), a selector, and a retraction. Each retraction is given as a map that is \(C^2\) on some neighborhood of \(M\) and is the identity on \(M\), extended by a fixed point of \(M\) outside the neighborhood as in Eq.~\eqref{eq:retraction}. In each case, the decoder applies the channel \(F(x)\) for the label \(x\in\Gamma_h\). By Proposition~\ref{prop:rounding} and Lemma~\ref{lem:retraction}, the construction in the main text yields a program of cost \(\frac\nu2\log(1/\eps)+O(1)\) from \(O(\eps^{-1/2})\) queries.

\paragraph*{\(\PU(d)\).}
Let \(\nu=d^2-1\), fix a Hilbert--Schmidt orthonormal basis \(E_1,\ldots,E_\nu\) of the traceless Hermitian matrices, and put
\[
  B:=[-\pi\sqrt d,\pi\sqrt d]^\nu,
  \qquad
  F(t):=\cU_{\exp(i\sum_at_aE_a)} .
\]
The selector chooses the representative \(U\) of \(u\) whose entry of largest absolute value (the lexicographically first one in case of ties) is a positive real number, forms the matrix logarithm \(A:=\Arg(U)\) with the arguments of the eigenvalues taken in \((-\pi,\pi]\), and puts \(t_a:=\Tr(E_aA)\). Since \(\sum_at_aE_a=A-\frac{\Tr A}{d}I\), we have \(F(t)=\cU_u\), and \(\norm t_2\le\norm A_2\le\pi\sqrt d\) gives \(t\in B\). Since the family is the whole of \(\PU(d)\), we take \(\Pi=\id\). Lemma~\ref{lem:retraction} then reduces to Eq.~\eqref{eq:haah-bias}, and the number of queries is \(O(d^2\eps^{-1/2})\).

\paragraph*{Phase gates.}
The family \(M_{\mathrm{ph}}:=\{[U_\theta]\}_{\theta\in[0,2\pi)}\subset\PU(2)\) of the qubit phase gates \(U_\theta:=\Ket0\!\Bra0+e^{-i\theta}\Ket1\!\Bra1\) is a circle, and \(\nu=1\). We put
\[
  B_{\mathrm{ph}}:=[0,2\pi],
  \qquad
  F_{\mathrm{ph}}(\theta):=\cU_{U_\theta},
\]
and the selector chooses the representative \(U\) of \(u\) with \(U_{00}=1\) and returns \(\theta\in[0,2\pi)\) with \(U_{11}=e^{-i\theta}\). The retraction replaces each diagonal entry of a representative \(\widehat U\) of the estimate by its phase:
\[
  \Pi_{\mathrm{ph}}(\widehat u)
  :=\Bigl[\sum_{k=0}^1\frac{\widehat U_{kk}}{\lvert\widehat U_{kk}\rvert}\Ket k\!\Bra k\Bigr].
\]
The right-hand side does not depend on the phase of the representative, is smooth on the neighborhood of \(M_{\mathrm{ph}}\) where \(\widehat U_{00}\widehat U_{11}\ne0\), and is the identity on \(M_{\mathrm{ph}}\).

\paragraph*{Reflections.}
For a unit vector \(\psi\in\C^d\), the reflection \(R_\psi:=I-2\Ket\psi\!\Bra\psi\) does not depend on the global phase of \(\psi\). The family \(M_{\mathrm{ref}}:=\{[R_\psi]\}\subset\PU(d)\) satisfies (R1), and \(\nu=2(d-1)\). Choose the global phase of \(\psi\) so that its first nonzero component is positive, and write \(\psi_k=r_ke^{i\phi_k}\) (\(\phi_1=0\)). We express the amplitudes by angles \(\theta\in[0,\pi/2]^{d-1}\) as
\[
  r_k=\cos\theta_k\prod_{j<k}\sin\theta_j\quad(1\le k\le d-1),
  \qquad
  r_d=\prod_{j<d}\sin\theta_j,
\]
and with the coordinates \(t:=(\theta,\phi_2,\ldots,\phi_d)\) we put
\[
  B_{\mathrm{ref}}:=[0,\pi/2]^{d-1}\times[-\pi,\pi]^{d-1},
  \qquad
  F_{\mathrm{ref}}(t):=\cU_{R_{\psi(t)}} .
\]
The phase of a component with zero amplitude and the angles following a \(\theta_j\) with \(\sin\theta_j=0\) are undetermined, so \(F_{\mathrm{ref}}\) is not injective; the selector sets the undetermined angles to \(0\).

The retraction is given by an eigendecomposition. The eigenvalues of \(R_\psi\) are \(-1\) (with eigenvector \(\psi\)) and \(1\) (with multiplicity \(d-1\)), and changing the phase of the representative multiplies all eigenvalues by the same phase. We therefore let \(\widehat\psi\) be an eigenvector for the eigenvalue \(\lambda_k\) maximizing \(\min_{j\ne k}\abs{\lambda_k-\lambda_j}\) among the eigenvalues \(\lambda_1,\ldots,\lambda_d\) of a representative \(\widehat U\) of the estimate, and put \(\Pi_{\mathrm{ref}}(\widehat u):=[R_{\widehat\psi}]\). On a neighborhood of \(M_{\mathrm{ref}}\), this eigenvalue is unique and simple for \(d\ge3\), and for \(d=2\) either choice gives the same point since \(R_{\psi^\perp}=-R_\psi\). Hence \(\Pi_{\mathrm{ref}}\) is smooth on a neighborhood of \(M_{\mathrm{ref}}\) and is the identity on \(M_{\mathrm{ref}}\).

\section{Computational complexity}
\label{sec:complexity}

The decoder reads the label \(x\) and applies the unitary \(V_x\). Every \(d\)-dimensional unitary decomposes into at most \(d(d-1)/2\) two-level unitaries~\cite{NielsenChuangBook}, and synthesizing each factor to accuracy \(\eps/O(d^2)\) by the Solovay--Kitaev theorem~\cite{Kitaev1997,DawsonNielsen2006} gives \(\operatorname{poly}(d)\cdot\operatorname{polylog}(1/\eps)\) gates.

The encoder rounds a point \(v\in M\) of the family unbiasedly, where \(v=u\) in the known case and \(v=\Pi(\widehat u)\) in the unknown case. The geometry of the family enters only through the evaluation of \(\Pi\) and the computation of coordinates of \(v\). We bound the error when the rounding is carried out with finite precision. We write the bound on the first derivative as
\[
  M_1:=\max_\alpha\sup_{t\in B_\alpha}\sup_{\norm w_2=1}\norm{DF_{\alpha,t}[w]}_\diamond .
\]

\begin{smlemma}[Perturbation stability]
\label{lem:perturbation}
For \(v\in M\), suppose that the encoder uses, instead of the selector value \(s(v)\), a pair \((\alpha,\tilde t)\) with \(\tilde t\in B_\alpha\) satisfying
\begin{equation}
  \tfrac12\norm{F_\alpha(\tilde t)-\cU_v}_\diamond\le\delta .
  \label{eq:approx-coord}
\end{equation}
Suppose further that, instead of the fractional part \(r_j\) of each coordinate of \(\tilde t\), it rounds with \(\tilde r_j\in[0,1]\) satisfying \(\abs{\tilde r_j-r_j}\le\eta\). Then the rounding error satisfies
\[
  \tfrac12\bigl\lVert\E[F_\alpha(X)]-\cU_v\bigr\rVert_\diamond
  \le\frac{M_2\nu}{16}(1+4\eta^2)h^2
  +\frac{\sqrt\nu}{2}M_1\eta h
  +\delta .
\]
In particular, the error bound in Eq.~\eqref{eq:bary-error} holds with its rounding term \(\frac{M_2\nu}{16}h^2\) replaced by the right-hand side.
\end{smlemma}

Condition~\eqref{eq:approx-coord} does not require \(\tilde t\) to be close to the coordinates \(s(v)\). Near discontinuities of the selector, numerical errors may return coordinates in another rectangle or on another branch without affecting the bound.

\begin{proof}
By the triangle inequality, it suffices to bound \(\frac12\norm{\E[F_\alpha(X)]-F_\alpha(\tilde t)}_\diamond\). Since \(X_j\) takes the adjacent lattice points enclosing \(\tilde t_j\) with probabilities \(1-\tilde r_j\) and \(\tilde r_j\),
\[
  \E[X_j]-\tilde t_j=(\tilde r_j-r_j)\delta_{\alpha,j},
  \qquad
  \E[(X_j-\tilde t_j)^2]
  =\bigl(\tilde r_j(1-\tilde r_j)+(\tilde r_j-r_j)^2\bigr)\delta_{\alpha,j}^2
  \le\Bigl(\frac14+\eta^2\Bigr)h^2 .
\]
Hence \(\norm{\E[X]-\tilde t}_2\le\sqrt\nu\,\eta h\) and \(\E\norm{X-\tilde t}_2^2\le\nu(\frac14+\eta^2)h^2\). In the Taylor expansion around \(\tilde t\) as in the proof of Lemma~\ref{lem:taylor}, the first-order term no longer vanishes but contributes only \(\norm{DF_{\alpha,\tilde t}[\E[X]-\tilde t]}_\diamond\le M_1\sqrt\nu\,\eta h\), and the remainder is at most \(\frac12M_2\E\norm{X-\tilde t}_2^2\).
\end{proof}

When \(h=\Theta(\sqrt\eps)\), the error changes only by a constant factor if \(\delta=O(\eps)\) and \(\eta=O(\sqrt\eps)\). The fractional parts may be computed less precisely than the coordinates because an error in the fractional parts shifts the mean of the rounding only by \(\eta h\).

\begin{proposition}
\label{prop:encoder}
Suppose that an oracle is given that returns, for \(v\in M\), a pair \((\alpha,\tilde t)\) satisfying condition~\eqref{eq:approx-coord} with \(\delta=O(\eps)\). Then an encoder with error \(\eps\) can be realized with, in addition to the \(\operatorname{poly}(d,n)\) operations of the estimation stage and one call each to \(\Pi\) and to the oracle, \(O(\nu)\) arithmetic operations on \(\Theta(\log(1/\eps))\)-bit numbers and \(O(\nu\log(1/\eps))\) uniformly random bits. In the known case, the estimation stage and \(\Pi\) are unnecessary.
\end{proposition}

\begin{proof}
Rounding \(\tilde t\) within \(B_\alpha\) to \(\Theta(\log(1/\eps))\) bits changes \(F_\alpha(\tilde t)\) by at most \(M_1\) times \(O(\eps)\), so condition~\eqref{eq:approx-coord} remains satisfied with \(\delta=O(\eps)\). For each coordinate, compute the cell index and the fractional part \(r_j\), and let \(\tilde r_j\) be \(r_j\) truncated to \(b=\ceil{\frac12\log(1/\eps)}\) bits; then \(\eta\le2^{-b}\le\sqrt\eps\). Drawing a uniform \(b\)-bit integer and comparing it with \(2^b\tilde r_j\) yields a bit that is \(1\) with probability exactly \(\tilde r_j\). This is done for each of the \(\nu\) coordinates. The label is a tuple of integers consisting of the cell indices and \(\alpha\), and the real values of the lattice points need not be computed. By Lemma~\ref{lem:perturbation}, the error changes only by a constant factor.
\end{proof}

The constants \(M_1\), \(M_2\), and those of Lemma~\ref{lem:retraction}, including the radius of the tubular neighborhood, are not necessarily computable uniformly from a finite description of \(M\). The mesh \(h\) and the number of queries \(n\) depend on them and are therefore chosen in a family-dependent preprocessing step.

\section{Proof of the lower bound}
\label{sec:lower}

\begin{smtheorem}
\label{thm:lower}
If \(M\) satisfies (R1), there is \(C<\infty\) depending only on \(M\) such that for all sufficiently small \(\eps>0\),
\[
  c_P^*(\eps;\cS)
  \ge
  \frac\nu2\log\frac1\eps
  -\frac{3\nu}{2}\log\log\frac1\eps
  -C .
\]
\end{smtheorem}

It suffices to prove the lower bound for a subfamily. We therefore take one local coordinate system of \(M\) and consider a \(C^1\) embedding \(t\mapsto u_t\in M\) of a compact rectangle \(B\subset\R^\nu\). We write \(V>0\) for the volume of \(B\).

\subsection{Heisenberg estimation of the coordinates}

\begin{smlemma}
\label{lem:bilipschitz}
There are \(a,b>0\) depending only on \(B\) and \(t\mapsto u_t\) such that
\[
  a\norm{s-t}_2
  \le d_F(u_s,u_t)
  \le b\norm{s-t}_2
  \qquad(s,t\in B).
\]
\end{smlemma}

\begin{proof}
Put \(G(t):=J(u_t)\). By Eq.~\eqref{eq:dF-choi}, \(d_F(u_s,u_t)=\norm{G(s)-G(t)}_2/\sqrt2\), so it suffices to prove the claim for \(G\). Since \(J\) is a smooth embedding, \(G\) is an injective \(C^1\) immersion.

Upper bound: with \(b':=\max_{t\in B}\norm{DG_t}<\infty\), convexity of \(B\) gives \(G(s)-G(t)=\int_0^1DG_{t+r(s-t)}[s-t]\,\dd r\), and hence \(\norm{G(s)-G(t)}_2\le b'\norm{s-t}_2\).

Lower bound: if there were no uniform lower bound, there would be distinct \(s_n,t_n\in B\) with
\[
  \frac{\norm{G(s_n)-G(t_n)}_2}{\norm{s_n-t_n}_2}\to0 .
\]
Passing to a subsequence, we may assume \(s_n\to s\) and \(t_n\to t\). If \(s\ne t\), this contradicts the injectivity of \(G\). If \(s=t\), then with \(v_n:=(s_n-t_n)/\norm{s_n-t_n}_2\to v\),
\[
  \frac{G(s_n)-G(t_n)}{\norm{s_n-t_n}_2}
  =\int_0^1DG_{t_n+r(s_n-t_n)}[v_n]\,\dd r
  \longrightarrow DG_s[v],
\]
so \(DG_s[v]=0\), contradicting that \(G\) is an immersion.
\end{proof}

Since the communication bound used below is formulated for parallel uses of the channel, we use the parallel estimator of Yang, Renner, and Chiribella (YRC)~\cite{YangRennerChiribella2020}.

\begin{smlemma}[Parallel estimator of YRC]
\label{lem:yrc-estimator}
There are \(C<\infty\) and \(n_d\ge1\) depending only on \(d\) such that for every \(n\ge n_d\) there exist a pure probe for \(n\) parallel queries to \(\cU_u\) and a covariant POVM that outputs an estimate \(\widehat u\in\PU(d)\). The distribution \(\mu_n\) of the relative error \(w=u^{-1}\widehat u\) does not depend on \(u\) and satisfies
\begin{equation}
  \int d_F(e,w)^2\,\mu_n(\dd w)\le\frac{C}{n^2}.
  \label{eq:yrc-mse}
\end{equation}
\end{smlemma}

\begin{proof}
We use the probe and the POVM of Ref.~\cite[Appendix~B, Eqs.~(B5)--(B8)]{YangRennerChiribella2020}. The conditional density has the form \(p_n(\widehat U\mid U)=\bigl\lvert\sum_\lambda\sqrt{q_{\lambda,n}}\,\chi_\lambda(U\widehat U^{-1})\bigr\rvert^2\), and since characters are class functions, the distribution of \(W\) in \(\widehat U=UW\) does not depend on \(U\). The loss \(1-d^{-2}\abs{\Tr W}^2\) used by YRC is exactly \(d_F(e,w)^2\), so the bound of Ref.~\cite[Theorem~2 and Eqs.~(B9), (B18)]{YangRennerChiribella2020} is Eq.~\eqref{eq:yrc-mse}.
\end{proof}

\begin{smlemma}
\label{lem:heisenberg}
There are \(c>0\) and \(n_0\ge1\) depending only on the family \(\{\cU_{u_t}\}_{t\in B}\) such that for every \(n\ge n_0\) there exists a strategy that outputs an estimate \(\widehat t\in B\) from \(n\) parallel queries and satisfies
\[
  \sup_{t\in B}\MSE_n(t)\le cn^{-2},
  \qquad
  \MSE_n(t):=\E_t\norm{\widehat t-t}_2^2 .
\]
\end{smlemma}

\begin{proof}
Map the output \(\widehat u\) of Lemma~\ref{lem:yrc-estimator} to the point \(\widehat t:=\operatorname{argmin}_{x\in\Lambda}d_F(u_x,\widehat u)\) of a lattice \(\Lambda\subset B\) with mesh \(n^{-2}\) (ties broken lexicographically). Taking \(s\in\Lambda\) with \(\norm{s-t}_2\le\sqrt\nu\,n^{-2}\), the triangle inequality and the minimality of \(\widehat t\) give
\[
  d_F(u_{\widehat t},u_t)
  \le d_F(u_{\widehat t},\widehat u)+d_F(\widehat u,u_t)
  \le 2d_F(\widehat u,u_t)+d_F(u_s,u_t).
\]
By Lemma~\ref{lem:bilipschitz}, \(\norm{\widehat t-t}_2\le\frac2a d_F(\widehat u,u_t)+\frac{b\sqrt\nu}{a}n^{-2}\). Squaring, taking expectations, and using Eq.~\eqref{eq:yrc-mse}, we obtain \(\MSE_n(t)\le 8C/(a^2n^2)+2b^2\nu/(a^2n^4)\).
\end{proof}

\subsection{Recycling and communication}

\begin{smlemma}[Recycling of YRC~\cite{YangRennerChiribella2020}]
\label{lem:recycling}
Let \((\cC,\{\sigma_t\}_{t\in B})\) be an \(\eps\)-processor for the family \(\{\cU_{u_t}\}_{t\in B}\). For every integer \(m\ge1\) there is a \(t\)-independent decoder \(\cR_m\) such that
\[
  \sup_{t\in B}
  \norm{\cR_m(\,\cdot\,\otimes\sigma_t)-\cU_{u_t}^{\otimes m}}_\diamond
  \le 4\sqrt2\,m\sqrt\eps .
\]
\end{smlemma}

The left-hand side is the diamond norm without the factor \(\frac12\). The construction follows Ref.~\cite[Appendix~A]{YangRennerChiribella2020} and does not assume that the program is pure.

\begin{smlemma}[Communication lower bound of Yang, Chiribella, and Hayashi~\cite{YangChiribellaHayashi2020}]
\label{lem:ych}
Suppose that the family \(\{\cU_{u_t}\}_{t\in B}\) has an estimator with \(n\) parallel queries satisfying \(\sup_t\MSE_n(t)\le cn^{-\beta}\). For \(\eta\ge0\) and \(p\in(0,1-\eta)\), suppose that the volume of the Euclidean ball of radius \(r:=\sqrt{c/((1-p-\eta)n^\beta)}\) is at most \(2^\nu pV\). Then every protocol that sends a state on \(\cH_P\) once and implements \(\cU_{u_t}^{\otimes n}\) within diamond error \(\eta\) satisfies
\begin{equation}
  \log\dim\cH_P
  \ge
  \frac{p\nu\beta}{2}\log n
  -\frac{p\nu}{2}\log\frac1{1-p-\eta}
  -C ,
  \label{eq:ych}
\end{equation}
where \(C\) depends only on \(c\), \(\nu\), and \(B\).
\end{smlemma}

\begin{proof}
This follows from Ref.~\cite[Theorem~1, Eq.~(34)]{YangChiribellaHayashi2020}.
\end{proof}

\subsection{Proof of Theorem~\ref{thm:lower}}

Increasing the number \(m\) of recyclings increases the first term of Eq.~\eqref{eq:ych}, but it also increases the accumulated error \(\eta\propto m\), which eventually violates \(1-p-\eta>0\). We therefore introduce a parameter \(\delta\) and set \(\eta\le\delta/2\) and \(p=1-\delta\).

\begin{proof}[Proof of Theorem~\ref{thm:lower}]
Fix an \(\eps\)-processor and restrict it to the subfamily given by the local coordinates above. Put
\[
  \ell:=\log\frac1\eps,
  \qquad
  \delta:=\frac1\ell,
  \qquad
  m:=\floor{\frac{\delta}{8\sqrt2\,\sqrt\eps}},
\]
and assume that \(\eps\) is small enough that \(\ell\ge2\) and \(m\ge n_0\). Below, \(O(1)\) depends only on \(c\), \(\nu\), and \(B\).

\textit{Recycling.}
By Lemma~\ref{lem:recycling}, a single copy of \(\sigma_t\) yields an implementation of \(\cU_{u_t}^{\otimes m}\) with error \(\eta:=4\sqrt2\,m\sqrt\eps\le\delta/2\). Since \(m=\Theta(\delta\eps^{-1/2})\) and \(m\le\eps^{-1/2}\),
\begin{equation}
  \log m=\frac\ell2-\log\ell+O(1),
  \qquad
  \log m\le\frac\ell2 .
  \label{eq:log-m}
\end{equation}

\textit{Communication lower bound.}
By Lemma~\ref{lem:heisenberg} there is an estimator with \(\beta=2\). We apply Lemma~\ref{lem:ych} with \(n=m\) and \(p=1-\delta\). Then \(1-p-\eta=\delta-\eta\ge\delta/2=1/(2\ell)>0\). Moreover, \(r^2\le 2c\ell/m^2=O(\eps\ell^3)\to0\), so the condition on the volume of the ball is also satisfied.

\textit{Evaluation.}
Substituting \(\delta\log m\le1/2\) and \(1-p-\eta\ge1/(2\ell)\) into Eq.~\eqref{eq:ych} gives
\[
  \log\dim\cH_P
  \ge\nu(1-\delta)\log m-\frac\nu2\log(2\ell)-O(1)
  \ge\nu\log m-\frac\nu2\log\ell-O(1).
\]
Substituting Eq.~\eqref{eq:log-m}, we obtain \(\log\dim\cH_P\ge\frac\nu2\ell-\frac{3\nu}2\log\ell-O(1)\).
\end{proof}

The correction term \(\frac{3\nu}{2}\log\ell\) breaks down as follows. Reducing the number of recyclings to \(\delta\) times \(\eps^{-1/2}\) loses \(\nu\log(1/\delta)\), and setting \(1-p=\delta\) loses \(\frac\nu2\log(1/\delta)\). Balancing these against the loss \(\frac{\nu\delta}{2}\ell\) caused by \(p<1\) shows that \(\delta=\Theta(1/\ell)\) is optimal. Removing this term requires a sharper bound on the accumulated recycling error \(O(m\sqrt\eps)\).

\end{document}